\documentclass[]{tokenwave}

\usepackage[utf8]{inputenc}
\usepackage{url}
\usepackage{amsfonts}
\usepackage{pifont}
\usepackage{amsmath}
\usepackage{nicefrac}
\usepackage{wrapfig}
\usepackage{tikz}
\usepackage{xspace}
\usepackage{array}
\usepackage{adjustbox}
\usepackage{longtable}
\usepackage{placeins}
\usepackage{float}
\usepackage{fontawesome}
\usetikzlibrary{positioning, arrows.meta, fit, backgrounds, calc}

\tcbuselibrary{listings}
\newtcblisting{promptbox}[2][]{%
  enhanced,
  colback=white,
  colframe=twpurple,
  colbacktitle=twpurple,
  coltitle=white,
  width=\linewidth,
  arc=2mm,
  boxrule=0.5mm,
  title={\normalsize\faInfoCircle\hspace{0.5em}#2},
  breakable=true,
  fonttitle=\bfseries\Large,
  listing only,
  listing options={
    language={},
    basicstyle=\ttfamily\footnotesize,
    columns=fullflexible,
    keepspaces=true,
    showstringspaces=false,
    breaklines=true,
    breakatwhitespace=false,
    breakindent=1em,
    tabsize=2
  },
  drop shadow southeast,
  left=5pt,right=5pt,
  top=2pt,bottom=2pt,
  box align=top,
  before=\par\noindent,
  after=\par\vspace{3pt},
  #1
}

\definecolor{casegap}{HTML}{A93C58}
\definecolor{casediscovery}{HTML}{16756C}
\definecolor{caseboundary}{HTML}{506AA3}
\definecolor{caseastrweak}{HTML}{B26A00}
\definecolor{caseastrasuccess}{HTML}{2E7D32}
\definecolor{caseastraassert}{HTML}{7B3F98}
\newtcolorbox{trajectorycase}[3][]{%
  enhanced,breakable,
  colback=white,colframe=#2,colbacktitle=#2,coltitle=white,
  width=\linewidth,arc=2mm,boxrule=0.45mm,
  title={#3},fonttitle=\sffamily\bfseries\normalsize,
  fontupper=\small,
  left=7pt,right=7pt,top=5pt,bottom=5pt,
  before skip=8pt,after skip=8pt,
  before upper={\setlength{\parskip}{4pt}},
  #1
}
\lstdefinestyle{trajectoryexcerpt}{%
  language={},basicstyle=\ttfamily\fontsize{8.5pt}{10pt}\selectfont,
  columns=fullflexible,keepspaces=true,showstringspaces=false,
  breaklines=true,breakatwhitespace=true,breakindent=1em,
  frame=none,aboveskip=3pt,belowskip=3pt,tabsize=2
}

\definecolor{dividergray}{RGB}{240,240,240}
\definecolor{ourslavender}{RGB}{239,237,255}

\definecolor{resultfirst}{HTML}{9FBFDE}
\definecolor{resultsecond}{HTML}{C5D8EB}
\definecolor{resultthird}{HTML}{E3ECF5}
\definecolor{resultborder}{HTML}{6F9AC9}
\newcommand{\resultchip}[3]{%
  \tikz[baseline=(chip.base)]\node[rounded corners=1.6pt, inner xsep=0pt, inner ysep=1.1pt,
    minimum width=2.75em, left color=#1, right color=#1!30!white, draw=#2, line width=0.45pt]
    (chip) {#3};}
\newcommand{\resultfirst}[1]{\resultchip{resultfirst}{resultborder}{\textbf{#1}}}
\newcommand{\resultsecond}[1]{\resultchip{resultsecond}{none}{#1}}
\newcommand{\resultthird}[1]{\resultchip{resultthird}{none}{#1}}
\newcommand{\benchyes}{\textcolor{green!45!black}{\ding{51}}}
\newcommand{\benchno}{\textcolor{red!70!black}{\ding{55}}}

\newcommand{\modelicon}[1]{%
  \makebox[1.15em][c]{\raisebox{-0.18em}{\includegraphics[height=0.95em]{figures/#1}}}\hspace{0.25em}}
\newcommand{\iconkimi}{\modelicon{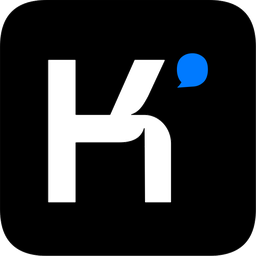}}
\newcommand{\icondeepseek}{\modelicon{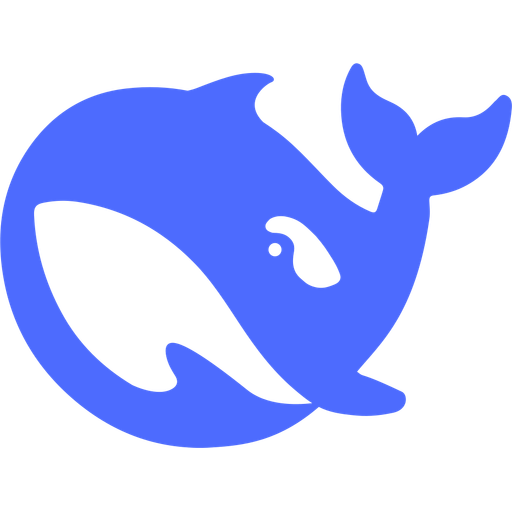}}
\newcommand{\iconglm}{\modelicon{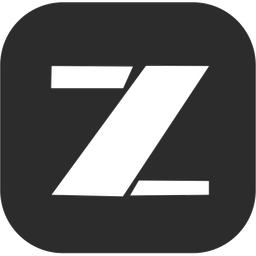}}
\newcommand{\iconclaude}{\modelicon{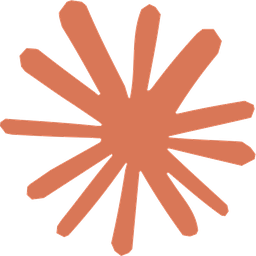}}
\newcommand{\icongpt}{\modelicon{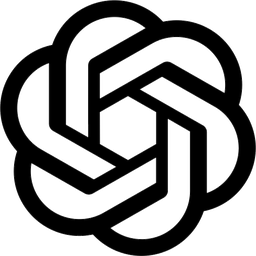}}

\crefname{figure}{Figure}{Figures}
\Crefname{figure}{Figure}{Figures}
\crefname{table}{Table}{Tables}
\Crefname{table}{Table}{Tables}
\crefname{section}{Section}{Sections}
\Crefname{section}{Section}{Sections}
\crefname{subsection}{Section}{Sections}
\Crefname{subsection}{Section}{Sections}
\crefname{equation}{Equation}{Equations}
\Crefname{equation}{Equation}{Equations}

\titleformat*{\paragraph}{\sffamily\bfseries}

\title{SpecEngineeringBench: Evaluating Coding Agents as Autonomous Specification Engineers}
\hypersetup{pdftitle={SpecEngineeringBench: Evaluating Coding Agents as Autonomous Specification Engineers}}

\author[1]{Xingjian Wang}
\author[2]{Shijian Wang}
\author[3]{Minghao Liu}
\author[2]{Ji Zou}
\author[1]{Zetao Yang}
\author[1]{Xinping Lei}
\author[1]{Yiming Liang}
\author[2]{Shen Yan}
\author[4]{Enduo Zhao}
\author[1]{Xiaoyang Shen}
\author[1,3]{Weihao Xuan}
\author[2]{Jiaheng Liu}
\author[1]{Ge Zhang}

\affiliation[1]{TokenWave AI\protect\textsuperscript{\protect\textdagger}}
\affiliation[2]{Independent Researcher}
\affiliation[3]{The University of Tokyo}
\affiliation[4]{Monolith Management, US}

\abstract{%
Autonomous specification engineering requires coding agents to translate functional requirements into executable tests and use the results to discover and validate defects. However, existing benchmarks evaluate coding agents on explicitly specified test generation tasks, commonly providing an issue description, a designated test entry point, or files to inspect, leaving the ability to determine what to test and where to investigate insufficiently evaluated. Relying on such clues bypasses the challenge of mapping requirements to implementations and proactively uncovering unknown defects. We propose \textbf{SpecEngineeringBench}, comprising 100 instances from 78 open-source Python repositories across 16 application categories, with multiple recorded defects in each instance. Two independent tasks evaluate complementary capabilities: generating test suites and discovering defects through execution analysis. The evaluation combines test validity, defect detection, and coverage with localization precision and recall. Across seven frontier coding agents, results reveal that test validity and coverage decouple from defect exposure: agents achieve a 91.47\% Head Pass Rate and 80.48\% Change-focused Coverage alongside a 26.36\% Fail-to-Pass Rate, while localization recall drops from 78.11\% at the file level to 51.29\% at the function level. Autonomous test generation improves file, function, and hunk localization recall by 4.99, 2.77, and 2.69 percentage points over relying solely on existing repository test suites.
}

\correspondence{\email{Themaoqiu@gmail.com}}
\date{\today}
\metadata[Website]{\url{https://tokenwave-ai.github.io/SpecEngineeringBench-Page/}}
\metadata[Code]{\url{https://github.com/TokenWave-AI/SpecEngineeringBench}}
\metadata[Dataset]{\url{https://huggingface.co/datasets/Tokenwave-AI/SpecEngineeringBench}}

\begin{document}

\maketitle
\begingroup\renewcommand{\thefootnote}{\textdagger}\footnotetext{Located in Palo Alto, United States.}\endgroup

\section{Introduction}
Coding agents driven by large language models (LLMs) increasingly perform complex software engineering tasks~\citep{sweagent,openhands,rethinkingtests}. As agents write and modify production code, specification engineering verifies whether implementations satisfy intended behavioral requirements~\citep{testdrivensurvey,autocover}. In realistic software development, assessing a prerelease candidate is fundamentally an exploratory, multi-defect investigation: specification engineers are not handed localized bug reports or specific failure traces; rather, they must independently interpret high-level product specifications~\citep{logichunter,testdrivensurvey}, navigate codebases, author discriminating test cases, and write reproducible defect reports. This setting motivates our central research question:

\begin{quote}
\itshape
Given only functional specifications and an uninspected repository, can coding agents autonomously uncover, localize, and substantiate software defects with reproducible evidence?
\end{quote}

\begin{figure}[t]
    \centering
    \includegraphics[width=\textwidth]{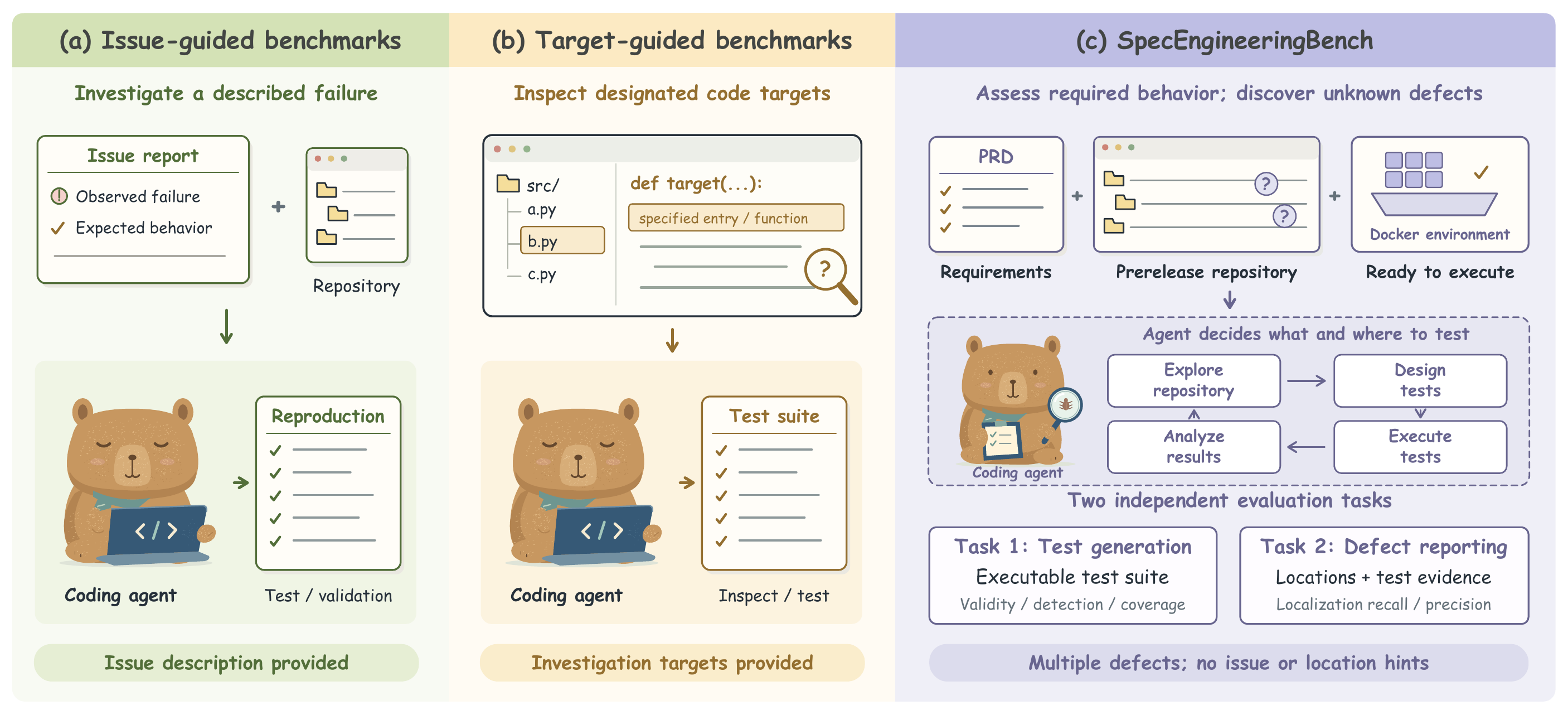}
    \caption{Conceptual comparison of representative benchmark settings and \textbf{SpecEngineeringBench}. \textbf{(a)} Issue-guided tasks provide a description of the faulty behavior~\citep{swtbench,tddbench}. \textbf{(b)} Target-guided tasks specify investigation targets, such as files, functions, or entry interfaces~\citep{testgeneval,testexplora}. \textbf{(c)} SpecEngineeringBench provides a PRD, a prerelease repository, and a configured environment, leaving agents to determine what and where to test. Agents explore the repository, design and execute tests, and analyze the results to investigate multiple unknown defects. Two independent tasks evaluate test-suite quality and defect reporting.}
    \label{fig:conceptual_comparison}
\end{figure}

Existing benchmarks address this question under two structural constraints (Figure~\ref{fig:conceptual_comparison}): (1) \textit{Investigation scopes assigned in advance:} Unit test generation and proactive defect search suites evaluate designated functions, entry interfaces, or target source files~\citep{testeval,testbench,testgeneval,testexplora,activeswe}. Fixing targets in advance evaluates localized code completion and omits the process of mapping functional requirements to production implementations. (2) \textit{Reliance on issue reports and single bug assumptions:} Benchmarks driven by issue reports supply descriptions of faulty behaviors or stack traces~\citep{swebench,swtbench,tddbench}, while reproduction suites verify isolated, known bugs~\citep{reproagent}. Prerelease software testing involves multiple unknown defects within the same repository, requiring agents to determine which behavioral surfaces require verification.

To bridge this gap, we argue that evaluating coding agents as autonomous specification engineers requires removing location hints and issue descriptions, requiring agents to derive test scopes and defect oracles from specifications. We introduce \textbf{SpecEngineeringBench}, a benchmark of 100 instances curated from 78 open-source Python repositories across 16 application categories, embedding 531 real-world defects (4--15 defects per instance). \textbf{SpecEngineeringBench} introduces four properties: \textbf{Requirements-based exploration:} Agents receive only a Product Requirements Document (PRD) specifying functional capabilities, boundary conditions, and error invariants, requiring them to determine what functionality to test. \textbf{Unrestricted repository scope:} The benchmark supplies no designated entry points or target files, evaluating exploration across the repository. \textbf{Multiple coexisting defects:} Each instance contains multiple defects, evaluating whether agents uncover coexisting issues across the codebase. \textbf{Dual tasks:} \textit{Task~1 (Test Generation)} evaluates whether an agent authors executable test suites; \textit{Task~2 (Defect Discovery and Reporting)} evaluates whether an agent executes tests, isolates root causes, and substantiates findings with reproducible evidence.

Evaluating seven frontier coding agents~\citep{claudeopus5, claudesonnet5, gpt56sol, gpt6astra, glm53, deepseekv41flash, kimik3} across these tasks yields three findings on autonomous defect discovery: (1) \textit{Passing tests and broad coverage do not imply defect discrimination:} Agents achieve high Head Pass Rates (91.47\%) and Change-focused Coverage (80.48\%), while the Fail-to-Pass Rate is 26.36\%. In 67.44\% of test outcomes, tests execute across both defective and repaired snapshots without detecting defects, as agents encode existing implementation outputs into assertions, treating the implementation rather than the specification as the source of truth, and omit discriminating oracles derived from the PRD. (2) \textit{Localization recall declines across structural granularities:} Agents attain 78.11\% recall at the file level, while recall drops to 51.29\% at the function level and 51.88\% at the hunk level. For defects spanning multiple files, recall drops to 2.00\%--7.84\%, identifying multi-file causal tracing as a recurring limitation. (3) \textit{Autonomous test generation increases defect recovery:} Agent-authored tests increase file, function, and hunk localization recall by 4.99, 2.77, and 2.69 percentage points over the baseline of executing existing repository test suites, showing the effect of active test authoring on defect identification.

\section{SpecEngineeringBench}
\label{sec:method}
\textbf{SpecEngineeringBench} evaluates coding agents under the conditions of specification conformance verification. Agents derive test scopes and defect oracles from product specifications without target hints or issue reports. The benchmark defines two complementary tasks: \textit{Task~1 (Test Generation)} measures whether an agent authors valid and discriminating test suites from specifications; \textit{Task~2 (Defect Discovery and Reporting)} measures whether executing and analyzing tests enables an agent to trace failures to root causes. Table~\ref{tab:benchmark_comparison} compares this benchmark with existing testing and repair suites.

\begin{table}[t]
\caption{Comparison of testing and software repair benchmarks. \textbf{Exploration scope}: designated search or test scope, even when broader repository context is accessible. \textbf{Issue-free}: no issue report is supplied. \textbf{Location-hint-free}: the task guarantees no designated investigation files, functions, or entry points; issue-driven tasks may carry such clues. \textbf{Test generation}: an evaluated task or track requires agents to author or complete tests. \textbf{Bug location}: defect localization is explicitly scored. \textbf{Multi-bug}: an instance explicitly combines multiple recorded defects. \textbf{Executable evidence}: agents are required to provide the evidence needed to reproduce the defects they report; reference tests used only to grade repairs do not qualify. \benchyes{} indicates that the property is present and \benchno{} indicates that it is absent. Independent replay of submitted report evidence is not implied.}
\label{tab:benchmark_comparison}
\centering
\fontsize{8pt}{10pt}\selectfont
\setlength{\tabcolsep}{3pt}
\renewcommand{\arraystretch}{1.17}
\begin{adjustbox}{max width=\textwidth}
\begin{tabular}{llcccccc}
\toprule
\textbf{Benchmark} & \shortstack{\textbf{Exploration}\\\textbf{scope}} & \shortstack{\textbf{Issue-}\\\textbf{free}} & \shortstack{\textbf{Location-}\\\textbf{hint-free}} & \shortstack{\textbf{Test}\\\textbf{generation}} & \shortstack{\textbf{Bug}\\\textbf{location}} & \textbf{Multi-bug} & \shortstack{\textbf{Executable}\\\textbf{evidence}} \\
\midrule
\rowcolor{dividergray}
\multicolumn{8}{c}{\textbf{\textit{Test generation}}} \\
TestEval~\citep{testeval} & Function & \benchyes & \benchno & \benchyes & \benchno & \benchno & \benchno \\
TestBench~\citep{testbench} & Method / class & \benchyes & \benchno & \benchyes & \benchno & \benchno & \benchno \\
TestGenEval~\citep{testgeneval} & Source file & \benchyes & \benchno & \benchyes & \benchno & \benchno & \benchno \\
CLOVER~\citep{clover} & Specified targets & \benchyes & \benchno & \benchyes & \benchno & \benchno & \benchno \\
\midrule
\rowcolor{dividergray}
\multicolumn{8}{c}{\textbf{\textit{Issue-driven reproduction and repair}}} \\
SWE-bench~\citep{swebench} & Repository & \benchno & \benchno & \benchno & \benchno & \benchno & \benchno \\
SWT-Bench~\citep{swtbench} & Repository & \benchno & \benchno & \benchyes & \benchno & \benchno & \benchyes \\
TDD-Bench Verified~\citep{tddbench} & Repository & \benchno & \benchno & \benchyes & \benchno & \benchno & \benchyes \\
SWE-Cycle~\citep{swecycle} & Repository & \benchno & \benchno & \benchyes & \benchno & \benchno & \benchyes \\
\midrule
\rowcolor{dividergray}
\multicolumn{8}{c}{\textbf{\textit{Proactive bug discovery and repair}}} \\
TestExplora~\citep{testexplora} & Entry interfaces & \benchyes & \benchno & \benchyes & \benchno & \benchno & \benchyes \\
Active-SWE~\citep{activeswe} & Review files & \benchyes & \benchno & \benchyes & \benchyes & \benchyes & \benchyes \\
SWE-sweep~\citep{swesweep} & Repository & \benchyes & \benchyes & \benchno & \benchno & \benchyes & \benchno \\
\midrule
\rowcolor{ourslavender}
\textbf{SpecEngineeringBench (Ours)} & \textbf{Repository} & \benchyes & \benchyes & \benchyes & \benchyes & \benchyes & \benchyes \\
\bottomrule
\end{tabular}
\end{adjustbox}
\end{table}

\subsection{Task Formulation}
\label{sec:task_formulation}

Each benchmark instance is formalized as a tuple $(r_i, d_i, e_i, \mathcal{B}_i)$, where $r_i$ is the unrepaired repository snapshot, $d_i$ is the product requirements document, $e_i$ is the isolated Docker runtime, and $\mathcal{B}_i$ ($|\mathcal{B}_i| \ge 4$) represents the set of coexisting reference defects. The agent $\mathcal{A}$ has access only to $(r_i, d_i, e_i)$, with no issue descriptions or target location hints provided. To separate test synthesis ability from runtime diagnostic reasoning, we establish two independent tasks:

\paragraph{Task 1: Test Generation ($\pi_1$).}
The agent generates a standalone test suite:
\begin{equation}
    \mathcal{T}_i^{(1)} = \mathcal{A}(r_i, d_i, e_i; \pi_1).
\end{equation}
The evaluation harness executes $\mathcal{T}_i^{(1)}$ against both the defective snapshot $r_i$ and the ground truth repaired snapshot $r_i^{+}$ out of band, scoring execution validity (HP), defect discrimination rate (F2P), and functional coverage (EC, CFG).

\paragraph{Task 2: Defect Discovery and Reporting ($\pi_2$).}
The agent conducts active exploration by authoring and executing temporary test probes $\mathcal{T}_i^{(2)}$ on $r_i$, producing a structured diagnostic report:
\begin{equation}
    \bigl(\mathcal{T}_i^{(2)}, \mathcal{R}_i\bigr) = \mathcal{A}(r_i, d_i, e_i; \pi_2).
\end{equation}
The submission $\mathcal{R}_i$ must substantiate each suspected defect with reproducible execution logs and source-grounded locations, scored against reference defects $\mathcal{B}_i$ across file, function, and edit-hunk granularities.

\subsection{Evaluation Protocol}
\label{sec:evaluation}

The evaluation protocol assesses two capabilities: generating test suites that cover specified functionality and check correctness, and discovering defects to report exact source locations. We evaluate test validity, defect detection, and coverage in Task~1, and localization recall and precision in Task~2.

\paragraph{Test validity and defect detection.}
\label{sec:test_evaluation}
For instance $i$, let $\mathcal{T}_i$ be the registered submitted tests, $\mathcal{H}_i$ the subset passing on the repaired snapshot $r_i^{+}$, and $\mathcal{F}_i$ the subset failing on the defective snapshot $r_i$ and passing on $r_i^{+}$. We compute Head Pass Rate (HP) by pooling test counts across the $M$ scored instances. Fail-to-Pass Rate (F2P) averages the fraction of discriminating tests within each instance:
\begin{equation}
    \mathrm{HP}=\frac{\sum_{i=1}^{M}|\mathcal{H}_i|}{\sum_{i=1}^{M}|\mathcal{T}_i|},\qquad
    \mathrm{F2P}=\frac{1}{M}\sum_{i=1}^{M}\frac{|\mathcal{F}_i|}{|\mathcal{T}_i|}.
    \label{eq:test_outcomes}
\end{equation}
HP checks compatibility with the repaired implementation; F2P requires a failure that disappears after repair, confirmed by repeated execution on both snapshots. F2P measures the share of generated tests that expose a defect. We report $\mathrm{Num}=M^{-1}\sum_i|\mathcal{T}_i|$ as the mean submitted suite size.

\paragraph{Functional and change coverage.}
Entry Coverage (EC) measures how broadly tests exercise the target functionality, while Change-focused Coverage (CFG) measures whether they reach the executable lines changed by the reference fixes. Let $\mathcal{E}_i$ denote executable file--line pairs in the reference coverage ranges, $\Delta_i$ executable changed file--line pairs, and $\mathcal{C}_i$ the union covered by the submitted suite, all on $r_i^{+}$. We average the coverage ratios across $M$ evaluated instances with valid environments:
\begin{equation}
    \mathrm{EC}=\frac{1}{M}\sum_{i=1}^{M}
    \frac{|\mathcal{C}_i\cap\mathcal{E}_i|}{|\mathcal{E}_i|},\qquad
    \mathrm{CFG}=\frac{1}{M}\sum_{i=1}^{M}
    \frac{|\mathcal{C}_i\cap\Delta_i|}{|\Delta_i|}.
    \label{eq:test_coverage}
\end{equation}
The suite level union counts each covered line once, and the outer mean gives each instance equal weight. These coverage targets remain evaluator only; agents must identify the implementation from the requirements.

\paragraph{Defect localization.}
\label{sec:report_evaluation}
Localization Recall (LR) measures the proportion of reference defect locations recovered by a report, while Localization Precision (LP) penalizes unsupported locations. For granularity $g\in\{\mathrm{file},\mathrm{function},\mathrm{hunk}\}$, let $G_i^g$ and $P_i^g$ denote the reference and predicted location sets. With matched reference count $a_i^g$ and credited prediction count $b_i^g$, we compute
\begin{equation}
    \mathrm{LR}_i^g=\frac{a_i^g}{|G_i^g|},\qquad
    \mathrm{LP}_i^g=\frac{b_i^g}{|P_i^g|}.
    \label{eq:localization}
\end{equation}
Files match by normalized paths; functions additionally require matching names. For hunks, a reported line group matches a reference edit when sharing a file and containing source lines within $\tau=3$ lines of each other. Hunk matching uses defective snapshot coordinates, with source line anchors for insertion edits.

\paragraph{Distinct bug localization.}
We measure the percentage of evaluable reference bugs successfully localized across all scored instances. Each source pull request defines one bug occurrence. A bug is localized when the report matches all reference files, function symbols, and more than half of the reference edit hunks. Function symbols include \texttt{<module>} for edits outside functions. Let $\mathcal{B}_i$ denote bugs with evaluable reference mappings in scored instance $i$, and $z_{ib}$ indicate whether bug $b$ satisfies these three conditions. We compute
\begin{equation}
    \text{Bugs Localized (\%)}=100\times
    \frac{\sum_{i=1}^{M}\sum_{b\in\mathcal{B}_i}z_{ib}}
    {\sum_{i=1}^{M}|\mathcal{B}_i|}.
    \label{eq:bug_localization}
\end{equation}
We apply the same calculation to bugs involving a single file or multiple files, and report the percentage of scored instances in which every evaluable bug is localized. Our success criterion evaluates submitted defect locations.

\subsection{Dataset Characteristics}
\label{sec:dataset}

\textbf{SpecEngineeringBench} comprises 100 instances from 78 open-source Python repositories spanning 16 application categories, with 531 recorded defect occurrences (Table~\ref{tab:dataset_statistics}). Each instance contains 4--15 defects, with a median of 5 and a mean of 5.31. To describe the extent of multi-defect assessment, Table~\ref{tab:dataset_statistics} groups instances by recorded defect count: (1) \textbf{4--5 defects} (74 instances); (2) \textbf{6--9 defects} (24 instances); and (3) \textbf{10--15 defects} (2 instances).

\begin{table}[t]
\caption{SpecEngineeringBench statistics by recorded defects per instance. \textbf{Repositories}: distinct upstream projects within each group; a project can occur in multiple groups, so the total is deduplicated. \textbf{Occurrences}: recorded defects counted separately in each instance. \textbf{Source kLOC}: non-documentation source code in the actual defective snapshots, excluding comments and blank lines but retaining tests, dependencies, and generated code ($1\,\mathrm{kLOC}=1{,}000$ lines). Source kLOC, reference files, and reference edit hunks are shown as median (range); files and hunks are the evaluator's localization targets.}
\label{tab:dataset_statistics}
\centering
\fontsize{9pt}{11pt}\selectfont
\setlength{\tabcolsep}{3pt}
\renewcommand{\arraystretch}{1.2}
\begin{tabular*}{\textwidth}{@{\extracolsep{\fill}}*{7}{c}@{}}
\toprule
\textbf{Defects / instance} & \textbf{Instances} & \textbf{Repositories} & \textbf{Occurrences} & \textbf{Source kLOC} & \textbf{Ref.\ files} & \textbf{Ref.\ hunks} \\
\midrule
4--5 & 74 & 62 & 341 & 73.9 (1.9--1,966.1) & 5 (1--21) & 10.5 (2--71) \\
6--9 & 24 & 20 & 162 & 102.3 (5.6--1,798) & 7.5 (4--12) & 19 (5--63) \\
10--15 & 2 & 2 & 28 & 147.5 (11.9--283) & 8.5 (4--13) & 21 (14--28) \\
\midrule
\textbf{Total} & \textbf{100} & \textbf{78} & \textbf{531} & \textbf{80.4 (1.9--1,966.1)} & \textbf{5 (1--21)} & \textbf{11.5 (2--71)} \\
\bottomrule
\end{tabular*}
\end{table}

\begin{figure}[t]
\centering
\includegraphics[width=\textwidth]{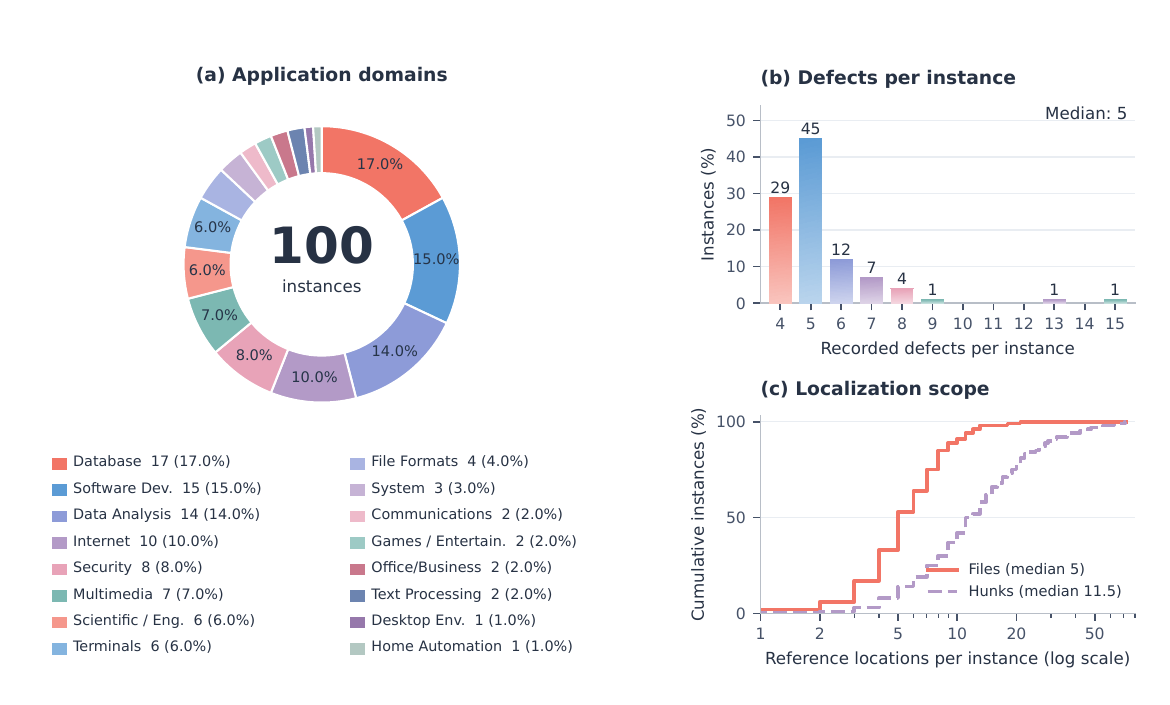}
\caption{Dataset composition and localization scope across all 100 instances. (a) Application-domain shares; the percentages and counts sum to 100 instances, drawn from 78 distinct repositories. (b) The distribution of recorded defects per instance; because there are 100 instances, bar counts equal percentages. (c) Empirical cumulative distributions across instances for the number of reference files (solid line) and reference edit hunks (dashed line); the horizontal axis is logarithmic and the vertical axis is the percentage of instances at or below each scope. Defect occurrences are counted separately in each instance.}
\label{fig:dataset_statistics}
\end{figure}

\paragraph{Repository scope and requirements.}
The 78 repositories represent 16 application categories under the repository inventory's classification. Figure~\ref{fig:dataset_statistics} shows the instance distribution across these categories, instance-level defect counts, and localization scopes. Their defective snapshots have a median of 80,449.5 non-documentation source lines (Table~\ref{tab:dataset_statistics}). Each instance pairs its snapshot with a product requirements document (PRD) describing the specified capabilities, expected results, boundary conditions, and error behavior, together with a Docker execution environment. The PRD specifies what should work; the agent must determine where and how to test it. Submitted suites distinguish unit, integration, and acceptance tests to exercise components, interactions, and observable behavior.

\paragraph{Defect scope and reference annotations.}
The recorded defects span a median of 5 reference files and 11.5 edit hunks per instance, requiring reports to identify multiple locations within the same repository. Our analysis covers 17 observed defect types in three families: \textbf{incomplete feature implementation} (5 types), \textbf{incorrect feature implementation} (7 types), and \textbf{generic programming errors} (5 types). The reference metadata links each recorded defect to its source PR, fix, regression tests, and source locations, and supplies the repaired snapshot and executable coverage ranges used in Section~\ref{sec:evaluation}.

\section{Experiments}
We evaluate frontier coding agents on \textbf{SpecEngineeringBench} across five core research questions: \textbf{RQ1} assesses the quality of tests generated for the specified functionality and agents' ability to localize defects through testing; \textbf{RQ2} measures the proportion of distinct reference bugs localized within each instance; \textbf{RQ3} investigates whether agents are more effective at discovering certain defect types than others; \textbf{RQ4} examines the mechanisms behind model failures in test construction and defect localization; and \textbf{RQ5} compares defect localization using repository-provided and agent-authored test suites.

\subsection{Experimental Setup}
\label{sec:setup}

\paragraph{Foundation models and agent harnesses.} We evaluate frontier models across native production
harnesses: Anthropic Claude (Claude-Opus-5~\citep{claudeopus5}, Claude-Sonnet-5~\citep{claudesonnet5} via Claude Code~\citep{claudecode}), OpenAI GPT (GPT-6-Astra~\citep{gpt6astra}, GPT-5.6-Sol~\citep{gpt56sol} via Codex~\citep{codexcli}), Moonshot Kimi (Kimi-K3~\citep{kimik3} via Claude Code~\citep{claudecode}), DeepSeek (DeepSeek-V4.1-Flash~\citep{deepseekv41flash} via Claude Code), and Z.ai GLM (GLM-5.3~\citep{glm53} via Claude Code) and under vendor recommended production configurations at high reasoning effort.

\paragraph{Container execution and environment isolation.} Benchmark tasks are packaged and verified using the standardized Harbor (v0.23.0) containerized framework~\citep{harbor}, ensuring out-of-band verifier execution. Trials execute in isolated containers with outbound networking disabled.


\subsection{RQ1: Test Validity Decouples from Defect Detection}

\begin{table}[t]
\caption{Main results on SpecEngineeringBench. Task~1 evaluates test generation and Task~2 evaluates defect discovery and reporting; HP pools test counts across scored instances; all other percentages and Num are averaged across scored instances. HP is Head Pass Rate (the fraction of registered tests that pass on the repaired snapshot), F2P is Fail-to-Pass Rate (the fraction that fail on the defective snapshot and pass on the repaired snapshot), EC is Entry Coverage (the fraction of executable target-functionality lines covered), and CFG is Change-focused Coverage (the fraction of executable reference changed lines covered). LR and LP are localization recall and precision, computed over the repository, at file, function, and hunk granularity; hunk matches use a tolerance of $\tau=3$ lines (see Section~\ref{sec:report_evaluation}). Num is the mean number of submitted test cases; all other metric values are percentages. For each percentage metric, the top three distinct values are shaded from darkest (best, bold) to lightest; ties share a shade. Num reports suite size.}
\centering
\fontsize{8pt}{10pt}\selectfont
\setlength{\tabcolsep}{2.5pt}
\renewcommand{\arraystretch}{1.25}
\begin{adjustbox}{max width=\textwidth}
\begin{tabular}{@{} >{\raggedright\arraybackslash}p{3.4cm} l *{11}{c} @{}}
\toprule
\multirow{3}{*}{\textbf{Model}} & \multirow{3}{*}{\textbf{Harness}} & \multicolumn{5}{c}{\textbf{Task 1: Test Generation}} & \multicolumn{6}{c}{\textbf{Task 2: Defect Discovery and Reporting}} \\
\cmidrule(lr){3-7} \cmidrule(lr){8-13}
 & & \multirow{2}{*}{\textbf{HP}} & \multirow{2}{*}{\textbf{F2P}} & \multirow{2}{*}{\textbf{EC}} & \multirow{2}{*}{\textbf{CFG}} & \multirow{2}{*}{\textbf{Num}}
 & \multicolumn{2}{c}{\textbf{File}} & \multicolumn{2}{c}{\textbf{Function}} & \multicolumn{2}{c}{\textbf{Hunk}} \\
\cmidrule(lr){8-9} \cmidrule(lr){10-11} \cmidrule(lr){12-13}
 & & & & & & & \textbf{LR} & \textbf{LP} & \textbf{LR} & \textbf{LP} & \textbf{LR} & \textbf{LP} \\
\midrule
\rowcolor{dividergray}
\multicolumn{13}{c}{\textbf{\textit{Open-Source Models}}} \\
\icondeepseek DeepSeek-V4.1-Flash\newline {\fontsize{7pt}{8pt}\selectfont \citep{deepseekv41flash}} & Claude Code & 92.25 & \resultsecond{27.69} & 71.29 & 76.89 & 86.88 & 76.97 & \resultthird{76.09} & 48.26 & \resultsecond{55.15} & 51.36 & \resultsecond{71.20} \\
\iconkimi Kimi-K3\newline {\fontsize{7pt}{8pt}\selectfont \citep{kimik3}} & Claude Code & \resultsecond{93.48} & 25.00 & \resultthird{77.53} & 81.20 & 120.50 & \resultthird{80.64} & \resultsecond{76.26} & 51.60 & \resultthird{54.94} & 51.98 & 69.72 \\
\iconglm GLM-5.3\newline {\fontsize{7pt}{8pt}\selectfont \citep{glm53}} & Claude Code & \resultfirst{94.40} & 21.33 & 76.64 & 80.58 & 142.02 & 79.78 & 75.42 & 49.39 & 54.92 & 51.57 & 68.73 \\
\rowcolor{dividergray}
\multicolumn{13}{c}{\textbf{\textit{Closed-Source Models}}} \\
\iconclaude Claude-Sonnet-5\newline {\fontsize{7pt}{8pt}\selectfont \citep{claudesonnet5}} & Claude Code & 90.65 & 21.78 & 68.23 & 72.88 & 68.74 & 62.89 & 73.82 & 40.08 & 50.49 & 41.20 & 66.33 \\
\iconclaude Claude-Opus-5\newline {\fontsize{7pt}{8pt}\selectfont \citep{claudeopus5}} & Claude Code & \resultthird{93.16} & 27.07 & \resultsecond{79.49} & \resultsecond{83.39} & 197.24 & \resultsecond{80.80} & \resultfirst{78.85} & \resultsecond{56.55} & 54.63 & \resultsecond{55.11} & \resultfirst{71.90} \\
\icongpt GPT-5.6-Sol\newline {\fontsize{7pt}{8pt}\selectfont \citep{gpt56sol}} & Codex & 86.02 & \resultfirst{34.09} & 72.41 & \resultthird{81.41} & 32.91 & 79.80 & 75.26 & \resultthird{52.59} & \resultfirst{58.55} & \resultthird{52.22} & \resultthird{70.34} \\
\icongpt GPT-6-Astra\newline {\fontsize{7pt}{8pt}\selectfont \citep{gpt6astra}} & Codex & 90.37 & \resultthird{27.56} & \resultfirst{81.41} & \resultfirst{87.01} & 162.62 & \resultfirst{85.85} & 71.59 & \resultfirst{60.56} & 47.72 & \resultfirst{59.72} & 63.27 \\
\bottomrule
\end{tabular}
\end{adjustbox}
\label{tab:main}
\end{table}

\paragraph{Execution validity converts into defect exposure at low rates.}
Table~\ref{tab:main} shows a numerical gap between test validity and defect exposure: mean HP reaches 91.47\% across seven agents, while mean F2P is 26.36\%. In an inspection of 49,317 registered tests from 500 evaluations across five systems, 67.44\% pass on the repaired snapshot without failing on the defective snapshot, 24.17\% satisfy F2P, and 8.39\% fail on the repaired snapshot. The primary outcome is a test that executes successfully while leaving defective behavior unexposed. Across these evaluations, 5.60\% contain passing repaired snapshot tests with zero F2P tests. Tests must distinguish supported behavior from deviations present in the defective implementation. Cases~\hyperref[casegap]{1} and~\hyperref[caseboundary]{3} demonstrate this pattern: broad passing suites leave explicit options or boundary contracts unchecked (see also Astra Cases~\hyperref[caseastrweak]{4}--\hyperref[caseastraassert]{6}).

\begin{table}[t]
\centering
\caption{F2P by Change-focused Coverage (CFG) interval for all seven models, each evaluated on 100 instances. Each entry is the mean per-instance F2P within the indicated CFG interval; both CFG and F2P are expressed as percentages. Intervals include their lower endpoint and exclude their upper endpoint, except the final interval also includes 100. N/A denotes an empty interval. The lowest-coverage interval contains at most three instances per model and is left unshaded; in every other interval, the three highest distinct values are shaded from darkest (best, bold) to lightest, and ties share a shade.}
\label{tab:test_detection_analysis}
\fontsize{8pt}{10pt}\selectfont
\setlength{\tabcolsep}{8pt}
\renewcommand{\arraystretch}{1.2}
\begin{adjustbox}{max width=\textwidth}
\begin{tabular}{lccccc}
\toprule
\multirow{2}{*}{\textbf{Model}} & \multicolumn{5}{c}{\textbf{Change-focused Coverage}} \\
\cmidrule(lr){2-6}
 & \textbf{0--20} & \textbf{20--40} & \textbf{40--60} & \textbf{60--80} & \textbf{80--100} \\
\midrule
\rowcolor{dividergray}
\multicolumn{6}{c}{\textbf{\textit{F2P}}} \\
\icondeepseek DeepSeek-V4.1-Flash & -- & 19.19 & \resultfirst{32.15} & \resultthird{22.96} & \resultsecond{30.06} \\
\iconkimi Kimi-K3 & 0.00 & \resultfirst{32.54} & 21.68 & 19.78 & 26.75 \\
\iconglm GLM-5.3 & -- & 21.49 & 15.47 & 12.25 & 24.40 \\
\iconclaude Claude-Sonnet-5 & 1.41 & 24.32 & 16.15 & 19.81 & 24.24 \\
\iconclaude Claude-Opus-5 & 0.00 & \resultthird{29.80} & \resultsecond{23.49} & 18.42 & \resultthird{29.90} \\
\icongpt GPT-5.6-Sol & 2.00 & \resultsecond{32.05} & 18.80 & \resultfirst{32.98} & \resultfirst{37.82} \\
\icongpt GPT-6-Astra & 1.12 & 7.35 & \resultthird{22.26} & \resultsecond{26.52} & 29.77 \\
\bottomrule
\end{tabular}
\end{adjustbox}
\end{table}

\paragraph{Reaching patched code does not ensure behavioral discrimination.}
Across seven agents, EC averages 75.29\% and CFG averages 80.48\%, indicating broad execution of target functionality and reference fix regions. Table~\ref{tab:test_detection_analysis} groups each model's instances by CFG. Even in the 80--100\% coverage interval, mean F2P ranges from 24.24\% to 37.82\% across models. Executing most reference changed lines therefore does not ensure that submitted tests distinguish the defective and repaired snapshots. Tests can execute relevant lines without supplying defect-triggering inputs or checking the affected behavior. Cases~\hyperref[casegap]{1} and~\hyperref[casediscovery]{2} illustrate requirements-derived assertions exposing behavior left unchecked by repository tests, while Astra Case~\hyperref[caseastrasuccess]{5} provides a Task~1 example of assertions that distinguish the two snapshots.

\paragraph{Defect reporting loses recall at finer localization granularity.}
Task~2 measures agreement with reference defect locations across file, function, and hunk granularities. Mean Localization Recall (LR) drops by 26.81 percentage points from the file level (78.11\%) to the function level (51.29\%), and measures 51.88\% at the hunk level. Localization Precision (LP) drops by 21.56 percentage points from the file level (75.33\%) to the function level (53.77\%) and rebounds to 68.78\% at the hunk level under the $\tau=3$ line tolerance. Agents identify relevant source files; once isolating a defective function, they pinpoint the edit hunk with minimal recall loss. Between these stages, agents face difficulty isolating root cause functions among surrounding callers and helpers within identified files.

\subsection{RQ2: Agents Leave Substantial Gaps in Distinct Bug Localization}
\label{sec:distinct_bug_localization}

\begin{table}[t]
\centering
\caption{Distinct bug localization in Task~2. A bug is localized when all reference files and function (including \texttt{<module>}) are matched and hunk recall exceeds 50\%, with a tolerance of three lines. Bugs Localized is the number of localized reference bugs divided by the number of evaluable reference bugs across scored instances, with each bug occurrence weighted equally. Overall includes all evaluable bugs; Single-file bugs and Multi-file bugs apply the same calculation to bugs whose reference locations span one file or multiple files, respectively. In each column, the three highest distinct values are shaded from darkest (best, bold) to lightest.}
\label{tab:bug_localization}
\fontsize{8pt}{10pt}\selectfont
\setlength{\tabcolsep}{6pt}
\renewcommand{\arraystretch}{1.2}
\begin{adjustbox}{max width=\textwidth}
\begin{tabular}{@{}lccc@{}}
\toprule
\textbf{Model} & \textbf{Overall} & \textbf{Single-file bugs} & \textbf{Multi-file bugs} \\
\midrule
\rowcolor{dividergray}
\multicolumn{4}{c}{\textbf{\textit{Open-Source Models}}} \\
\icondeepseek DeepSeek-V4.1-Flash & 37.93 & 46.19 & \resultthird{3.92} \\
\iconkimi Kimi-K3 & 42.72 & 52.14 & \resultthird{3.92} \\
\iconglm GLM-5.3 & 42.34 & 51.90 & 2.94 \\
\rowcolor{dividergray}
\multicolumn{4}{c}{\textbf{\textit{Closed-Source Models}}} \\
\iconclaude Claude-Sonnet-5 & 33.14 & 40.63 & 2.00 \\
\iconclaude Claude-Opus-5 & \resultsecond{45.37} & \resultsecond{54.57} & \resultfirst{7.84} \\
\icongpt GPT-5.6-Sol & \resultthird{43.30} & \resultthird{52.38} & \resultsecond{5.88} \\
\icongpt GPT-6-Astra & \resultfirst{49.04} & \resultfirst{59.05} & \resultfirst{7.84} \\
\bottomrule
\end{tabular}
\end{adjustbox}
\end{table}

\paragraph{Agents localize less than half of reference bugs.}
Across evaluated frontier models, the percentage of reference bugs successfully localized ranges from 33.14\% to 49.04\% (Table~\ref{tab:bug_localization}). GPT-6-Astra attains 85.85\% recall at the file level (Table~\ref{tab:main}), while localizing 49.04\% of distinct reference bugs across scored instances. Identifying an affected file through test execution does not isolate the underlying defective functions and code modifications. Authoring tests that trigger visible runtime exceptions can expose repository files without covering the specific boundary inputs required to localize the full defect.

\paragraph{Defects spanning multiple files show lower localization.}
Localization rates depend on bug scope. Across models, agents localize 40.63\%--59.05\% of bugs located within a single file, while localization rates decrease to 2.00\%--7.84\% for bugs spanning multiple files (Table~\ref{tab:bug_localization}). Multi-file defects involve contract mismatches and state transformations across component boundaries. Agents primarily write unit tests for isolated functions, producing test suites that omit execution paths traversing multiple modules. When a test fails, the agent investigates the immediate exception frame, leaving callers and dependent components uninspected.

\paragraph{Distribution of localized defects across instances.}
In instance-level evaluations, agents localize every evaluable bug in 4.00\% to 11.00\% of instances. In 70.71\% to 88.89\% of instances, submissions contain a combination of localized and unlocalized defects. After localizing at least one bug in an instance, GPT-6-Astra leaves remaining coexisting bugs unlocalized in 83.00\% of instances. Exploration frequently halts after confirming an initial failure, leaving coexisting defects in the repository undetected.

\subsection{RQ3: Discovery Capabilities Skew Across Defect Types}

\begin{figure}[t]
\centering
\includegraphics[width=0.82\textwidth]{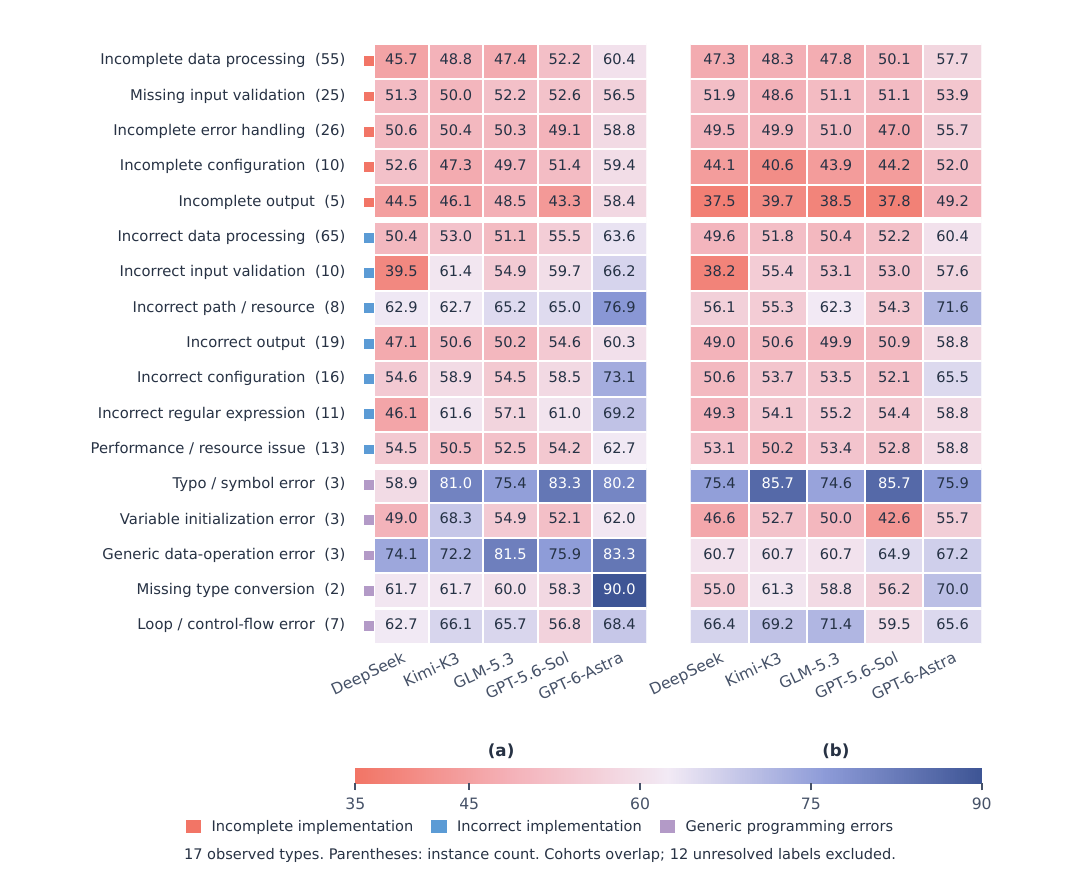}
\caption{Instance-averaged localization recall (\%) by defect-type cohort for the five evaluated systems with existing defect-type annotations. Panel (a) reports function-level recall and panel (b) reports hunk-level recall; columns within each panel are DeepSeek-V4.1-Flash, Kimi-K3, GLM-5.3, GPT-5.6-Sol, and GPT-6-Astra. Parentheses give the number of instances in each cohort; multiple defects of the same type within one instance do not increase its weight. These instances contain 442 recorded defect occurrences, of which 430 have type labels; 12 unresolved labels are excluded from cohort membership. The 14 replacement instances await type annotation.}
\label{fig:rq2_defect_types}
\end{figure}

\paragraph{Incomplete behaviors create shared blind spots across models.}
Figure~\ref{fig:rq2_defect_types} shows low recall across several recurring defect categories. In the \emph{Incomplete data processing} cohort (55 instances), function and hunk LR average 50.88\% and 50.23\% across the five models. The \emph{Missing input validation} cohort (25 instances) yields hunk LR between 48.64\% and 53.87\%. Output cohorts also show low recall, with mean hunk LR of 51.83\% for \emph{Incorrect output} (19 instances) and 40.56\% for \emph{Incomplete output} (five instances). Weaknesses recur across data transformations, validation routines, and output serialization, indicating that uncompleted behaviors challenge testing across all evaluated agents.

\paragraph{Localization accuracy within identified files varies across defect types.}
Conditioning on complete file recall isolates localization failures within known files. Across the five models, 46 model instance records in the \emph{Missing input validation} cohort identify every reference file; 95.65\% omit reference functions and 93.48\% omit reference hunks, with mean function and hunk LR at 66.51\% and 60.71\%. In the \emph{Incorrect configuration} cohort, 89.58\% and 91.67\% of the 48 records with complete file recall omit functions and hunks. In the \emph{Incorrect regular expression} cohort, omission rates drop to 44.00\% across 25 complete file records, where mean function LR reaches 90.57\%. Discovering all relevant files leaves differing residual gaps across defect types, with input validation and configuration requiring more specific inspection within discovered files.

\begin{table}[t]
\centering
\caption{Localization Recall (LR) and Localization Precision (LP) across defect-type cohorts for the five evaluated systems on the 86 annotated instances retained in the current benchmark. LR is the percentage of reference locations recovered, and LP is the percentage of reported locations credited by the evaluator, computed at file, function, and hunk granularity. Each entry is the mean over all eligible model--instance records in the cohort at the indicated granularity; cohorts overlap and each instance is evaluated against its complete reference set. In each column, the three cohorts with the highest values are shaded from darkest (bold) to lightest; ties share a shade.}
\label{tab:cohort_failure_analysis}
\fontsize{8pt}{10pt}\selectfont
\setlength{\tabcolsep}{5pt}
\renewcommand{\arraystretch}{1.2}
\begin{adjustbox}{max width=\textwidth}
\begin{tabular}{@{}lcccccc@{}}
\toprule
\multirow{2}{*}{\textbf{Defect type}} & \multicolumn{2}{c}{\textbf{File}} & \multicolumn{2}{c}{\textbf{Function}} & \multicolumn{2}{c}{\textbf{Hunk}} \\
\cmidrule(lr){2-3}\cmidrule(lr){4-5}\cmidrule(lr){6-7}
& \textbf{LR} & \textbf{LP} & \textbf{LR} & \textbf{LP} & \textbf{LR} & \textbf{LP} \\
\midrule
\rowcolor{dividergray}
\multicolumn{7}{c}{\textbf{\textit{Incomplete implementation}}} \\
Incomplete data processing & 76.83 & 72.03 & 50.88 & 50.58 & 50.23 & 65.70 \\
Missing input validation & 81.87 & 74.95 & 52.54 & 53.59 & 51.32 & 70.36 \\
Incomplete error handling & 76.51 & 73.38 & 51.85 & 53.88 & 50.62 & 68.62 \\
Incomplete configuration & 73.97 & 80.30 & 52.08 & 57.37 & 44.95 & 69.66 \\
Incomplete output & 71.89 & 76.13 & 48.17 & 59.59 & 40.56 & 65.26 \\
\rowcolor{dividergray}
\multicolumn{7}{c}{\textbf{\textit{Incorrect implementation}}} \\
Incorrect data processing & 79.04 & 75.09 & 54.71 & 54.28 & 52.89 & 67.95 \\
Incorrect input validation & 81.89 & 76.85 & 56.35 & 50.37 & 51.46 & 69.45 \\
Incorrect path / resource & 87.62 & 72.00 & \resultthird{66.52} & 48.97 & 59.95 & 66.81 \\
Incorrect output & 76.08 & 71.71 & 52.54 & 56.16 & 51.83 & 66.54 \\
Incorrect configuration & 85.65 & 76.06 & 59.89 & 49.60 & 55.08 & 67.65 \\
Incorrect regular expression & 76.12 & 79.55 & 59.00 & 58.41 & 54.38 & 71.53 \\
Performance / resource issue & 81.15 & \resultthird{84.40} & 54.87 & \resultthird{62.57} & 53.67 & \resultfirst{82.28} \\
\rowcolor{dividergray}
\multicolumn{7}{c}{\textbf{\textit{Generic programming errors}}} \\
Typo / symbol error & \resultthird{87.69} & 73.58 & \resultsecond{75.75} & 50.24 & \resultfirst{79.46} & 69.34 \\
Variable initialization error & 81.43 & \resultfirst{85.38} & 57.28 & \resultsecond{63.88} & 49.54 & \resultthird{76.01} \\
Generic data-operation error & \resultfirst{98.89} & \resultsecond{84.95} & \resultfirst{77.41} & \resultfirst{72.67} & \resultthird{62.84} & \resultsecond{80.66} \\
Missing type conversion & 80.00 & 66.36 & 66.33 & 48.73 & 60.25 & 58.02 \\
Loop / control-flow error & \resultsecond{94.40} & 73.82 & 63.92 & 62.32 & \resultsecond{66.45} & 70.28 \\
\bottomrule
\end{tabular}
\end{adjustbox}
\end{table}

\paragraph{Cohort recall metrics encompass omissions and uncredited predictions.}
The \emph{Performance / resource issue} and \emph{Incomplete data processing} cohorts yield similar mean hunk LR (53.67\% and 50.23\%), while their reports combine omissions and uncredited predictions differently (Table~\ref{tab:cohort_failure_analysis}). In the performance cohort, 35.38\% of reports attain complete hunk precision while omitting reference hunks, recovering 55.10\% of reference hunks on average. In the incomplete data processing cohort, 81.45\% of reports simultaneously omit reference hunks and include uncredited predictions, compared with 56.92\% in the performance cohort. Recovering these defects requires wider reference coverage alongside selective candidate reporting.

\subsection{RQ4: Behavioral Failure Modes in Test Construction and Localization}

\paragraph{Existing implementation behaviors replace specification-based oracles.}
Assertion mismatches account for 31.6\% of reviewed Task~1 failure mechanisms, and misconfigured mocks account for 37.5\%; invalid input or setup accounts for 7.4\%, API or parameter errors for 22.8\%, syntax or collection failures for 1.5\%, and runtime exceptions for 3.7\%. Inspection of execution trajectories indicates that agents infer test oracles from repository outputs, default fixtures, and existing return patterns, leaving requirements from the PRD unencoded in assertions. The resulting test suites execute cleanly while validating existing defective behaviors, failing to produce assertions that distinguish defective implementations from repaired ones.

\paragraph{Causal tracing halts at initial failure locations.}
Among reviewed localization failure mechanisms, premature search termination accounts for 36.0\%, evidence underuse for 30.9\%, and dependency truncation for 25.0\%. These failure modes reflect a shared pattern: upon observing a test failure or exception traceback, an agent anchors on the first visible error frame or entry point and halts exploration. Agents omit inspection of upstream caller hierarchies, cross file data flows, and execution logs needed to isolate the root cause. Submitted defect reports record immediate symptom locations while leaving the underlying defect path unexamined.

\paragraph{Defensive reporting trades completeness for precision.}
Reporting at an incorrect granularity accounts for 4.4\% of reviewed localization failure mechanisms, while conservative omissions account for 19.9\% and identifying symptoms as causes accounts for 25.0\%. Faced with causal ambiguity, agents employ two reporting strategies: they report coarse scopes by naming test drivers, entry interfaces, or configuration files in place of production code, or they omit plausible candidate locations to prevent uncredited predictions from reducing precision scores. This pruning produces sparse reports that omit confirmed defect locations and reduce the completeness of findings.

\subsection{RQ5: Autonomous Testing Improves Defect Recall Beyond Existing Suites}

We compare GPT-5.6-Sol with Codex on the same 100 instances. With \emph{Repository-provided tests}, the agent executes the existing test suite; with \emph{Agent-authored tests}, it authors tests from the PRD. Table~\ref{tab:rq4_testing_settings} compares localization recall and precision using the same paired reports and reference annotations, giving each instance equal weight.

\begin{table}[t]
\caption{Defect localization with repository-provided and agent-authored tests for GPT-5.6-Sol with Codex on the same 100 instances. LR and LP denote localization recall and precision, respectively. Values are instance-averaged percentages. Green upward and red downward annotations show changes in percentage points, computed before rounding.}
\label{tab:rq4_testing_settings}
\centering
\fontsize{8pt}{10pt}\selectfont
\setlength{\tabcolsep}{4pt}
\renewcommand{\arraystretch}{1.25}

\newcommand{\localizationscore}[3][green!45!black]{%
  \makebox[2.5em][r]{#2}%
  \makebox[3.4em][l]{\,\textcolor{#1}{\scriptsize #3}}%
}

\begin{adjustbox}{max width=\textwidth}
\begin{tabular}{lrrrrrr}
\toprule
\multirow{2}{*}{\textbf{Test suite}}
& \multicolumn{2}{c}{\textbf{File}}
& \multicolumn{2}{c}{\textbf{Function}}
& \multicolumn{2}{c}{\textbf{Hunk}} \\
\cmidrule(lr){2-3}
\cmidrule(lr){4-5}
\cmidrule(lr){6-7}
& \localizationscore{\textbf{LR}}{}
& \localizationscore{\textbf{LP}}{}
& \localizationscore{\textbf{LR}}{}
& \localizationscore{\textbf{LP}}{}
& \localizationscore{\textbf{LR}}{}
& \localizationscore{\textbf{LP}}{} \\
\midrule
Repository-provided tests
& \localizationscore{74.81}{}
& \localizationscore{78.00}{}
& \localizationscore{49.82}{}
& \localizationscore{60.16}{}
& \localizationscore{49.53}{}
& \localizationscore{73.59}{} \\
\rowcolor{green!8}
Agent-authored tests
& \localizationscore{79.80}{$\uparrow$+4.99}
& \localizationscore[red!75!black]{75.26}{$\downarrow$-2.74}
& \localizationscore{52.59}{$\uparrow$+2.77}
& \localizationscore[red!75!black]{58.55}{$\downarrow$-1.60}
& \localizationscore{52.22}{$\uparrow$+2.69}
& \localizationscore[red!75!black]{70.34}{$\downarrow$-3.25} \\
\bottomrule
\end{tabular}
\end{adjustbox}
\end{table}

\paragraph{Agent-authored testing increases average recall across a subset of instances.}
Table~\ref{tab:rq4_testing_settings} reports GPT-5.6-Sol results across 100 instances under two execution conditions: running repository-provided tests and authoring tests from the PRD. Authoring tests increases instance-averaged recall by 4.99 percentage points for files, 2.77 percentage points for functions, and 2.69 percentage points for hunks, while mean precision decreases by 2.74, 1.61, and 3.25 percentage points (differences computed before rounding). These recall gains concentrate in specific instances: function recall improves in 31\% of instances and hunk recall improves in 35\%, while the median paired change across all three granularities is zero. Authoring tests improves defect coverage in instances where additional tests support further discovery.

\paragraph{Recall gains expand partial discoveries through broader file identification.}
Among instances with improved function and hunk recall, 87.10\% and 91.43\% obtain nonzero recall under repository-provided tests, indicating that autonomous testing expands partial discoveries. Higher file recall accompanies 70.97\% of function improvements and 68.57\% of hunk improvements. Authoring tests broadens the set of affected files identified while recovering additional functions and hunks. Cases~\hyperref[casegap]{1}--\hyperref[caseboundary]{3} in Appendix~\ref{app:trajectory_case_studies} illustrate how the two testing settings produce different evidence for defect investigation.

\paragraph{Report growth exceeds matched locations and reduces precision.}
Agent-authored testing increases reported function locations from 868 to 1{,}011 (+16.47\%), while recovered reference functions increase from 521 to 572 (+9.79\%). Reported hunks increase from 527 to 608 (+15.37\%), while recovered reference hunks increase from 572 to 626 (+9.44\%). Total reported locations increase at a higher rate than confirmed defect sites, which lowers mean function and hunk precision by 1.60 and 3.25 percentage points (Table~\ref{tab:rq4_testing_settings}). Among instances with improved recall, 51.61\% at the function level and 40.00\% at the hunk level show lower precision. Agent-authored tests support broader discovery, while selecting candidate locations to report remains a constraint on precision.

\FloatBarrier

\section{Related Work}

\paragraph{Coding Agents.}
Agents based on large language models~\citep{sweagent,openhands,rethinkingtests} use tool environments to search repositories, edit source files, and execute commands over multiple turns. Existing benchmarks evaluate code generation~\citep{codetestbehavior}, repository level issue resolution~\citep{swebench}, complex feature implementation~\citep{featurebench}, and lifecycle workflows~\citep{swecycle}; SWE-smith~\citep{swesmith} expands training distributions for software engineering agents. Executing existing tests evaluates command execution without measuring an agent's ability to determine which behaviors require testing~\citep{tebench,rethinkingtests}. \textbf{SpecEngineeringBench} evaluates this test design capability under the configurations specified in Section~\ref{sec:setup}.

\paragraph{Benchmarks for Test Generation.}
Automated test generation includes regression testing~\citep{autocover}, search methods~\citep{llmsuite}, and iterative generation with execution feedback~\citep{testagent,coveredcode,autocover}. Existing benchmarks evaluate agents on specified targets: TestEval~\citep{testeval}, TestBench~\citep{testbench}, and TestGenEval~\citep{testgeneval} isolate individual functions, classes, and source files, while CLOVER~\citep{clover} examines coverage on specified components. Issue based benchmarks such as SWT-Bench~\citep{swtbench} and TDD-Bench Verified~\citep{tddbench} evaluate translating issue descriptions into reproduction tests~\citep{reproagent,swedoctor,echo}. In proactive defect discovery, TestExplora~\citep{testexplora} supplies entry interfaces, Active-SWE~\citep{activeswe} restricts inspection to review files, and SWE-sweep~\citep{swesweep} grades patches without evaluating executable test evidence. Coverage and execution metrics can overlook subtle defects~\citep{swebenchplus,covmutation,testoverfitting}. \textbf{SpecEngineeringBench} evaluates requirements-based testing on prerelease repositories with multiple coexisting defects.


\section{Conclusion}
We introduce \textbf{SpecEngineeringBench}, a benchmark for evaluating coding agents as autonomous specification engineers through requirements-driven testing of repositories containing multiple real-world defects. Its two independent tasks assess generated test quality and defect localization without supplying issue descriptions or investigation targets. Evaluations reveal a gap between test validity and defect detection: most submitted tests pass on repaired implementations, while substantially fewer expose defects. Higher coverage and larger test suites do not consistently translate into better defect detection or more precise localization, so each capability requires direct assessment. Relying solely on a repository's existing test suite reveals fewer defects than allowing agents to explore the repository and generate additional tests, demonstrating the value of active investigation beyond existing test coverage. As requirements are increasingly expanded into specifications automatically, verifying implementations against those specifications becomes the bottleneck of autonomous development. SpecEngineeringBench provides an empirical foundation for this verification capability, with future extensions to agent-generated specifications and additional programming languages.


\bibliographystyle{assets/plainnat}
\bibliography{references}

@misc{claudeopus5,
  title={Introducing {Claude Opus 5}},
  author={{Anthropic}},
  year={2026},
  month=jul,
  url={https://www.anthropic.com/news/claude-opus-5},
  note={Official model announcement, July 24, 2026}
}

@misc{claudesonnet5,
  title={Introducing {Claude Sonnet 5}},
  author={{Anthropic}},
  year={2026},
  month=jun,
  url={https://www.anthropic.com/news/claude-sonnet-5},
  note={Official model announcement, June 30, 2026}
}

@misc{gpt6astra,
  title={{GPT-6 Astra} Model},
  author={{OpenAI}},
  year={2026},
  url={https://developers.openai.com/api/docs/models/gpt-6-astra},
  note={Official model documentation. Accessed October 8, 2026}
}

@misc{gpt56sol,
  title={{GPT-5.6 Sol} Model},
  author={{OpenAI}},
  year={2026},
  url={https://developers.openai.com/api/docs/models/gpt-5.6-sol},
  note={Official model documentation. Accessed October 8, 2026}
}

@misc{kimik3,
  title={{Kimi K3} Is Here: Our Most Capable Model},
  author={{Moonshot AI}},
  year={2026},
  month=jul,
  url={https://forum.moonshot.ai/t/kimi-k3-is-here-our-most-capable-model/480},
  note={Official model announcement, July 22, 2026}
}

@misc{deepseekv41flash,
  title={{DeepSeek-V4.1-Flash} Model Card},
  author={{DeepSeek AI}},
  year={2026},
  url={https://huggingface.co/deepseek-ai/DeepSeek-V4.1-Flash},
  note={Official model card. Accessed October 15, 2026}
}

@misc{glm53,
  title={{GLM-5.3} Model Card},
  author={{Z.ai}},
  year={2026},
  url={https://huggingface.co/zai-org/GLM-5.3},
  note={Official model card. Accessed October 8, 2026}
}

@misc{claudecode,
  title={{Claude Code}: Overview},
  author={{Anthropic}},
  year={2026},
  url={https://code.claude.com/docs/en/overview},
  note={Official documentation. Accessed October 8, 2026}
}

@misc{codexcli,
  title={{Codex CLI}},
  author={{OpenAI}},
  year={2026},
  url={https://learn.chatgpt.com/docs/codex/cli},
  note={Official documentation. Accessed October 8, 2026}
}

@misc{harbor,
  title={{Harbor}: Task Format Overview},
  author={{Harbor Team}},
  year={2026},
  url={https://docs.harborframework.com/tasks/overview},
  note={Official documentation. Accessed October 8, 2026}
}

@article{rethinkingtests,
  title={Rethinking the Value of Agent-Generated Tests for LLM-Based Software Engineering Agents},
  author={Chen, Zhi and Sun, Zhensu and Shi, Yuling and Peng, Chao and Gu, Xiaodong and Lo, David and Jiang, Lingxiao},
  journal={arXiv preprint arXiv:2602.07900},
  year={2026}
}

@article{echo,
  title={Echo: Graph-Enhanced Retrieval and Execution Feedback for Issue Reproduction Test Generation},
  author={Fei, Zhiwei and Pan, Yue and Sarro, Federica and Ge, Jidong and Liu, Marc and Ng, Vincent and Ye, He},
  journal={arXiv preprint arXiv:2603.07326},
  year={2026}
}

@article{testagent,
  title={Multi-Agent LLM Collaboration for Unit Test Generation via Human-Testing-Inspired Workflows},
  author={Zhang, Quanjun and Shang, Ye and Gu, Siqi and Zhou, Jianyi and Fang, Chunrong and Chen, Zhenyu and Xiao, Liang},
  journal={arXiv preprint arXiv:2607.09101},
  year={2026}
}

@article{codetestbehavior,
  title={Do Code Language Models Follow Tests? Paired Interventions on Program Behavior},
  author={Liang, Yunhao and Gan, Chengguang and Ying, Ruixuan and Wei, Hanjun and Cui, Zhe and Ni, Shiwen},
  journal={arXiv preprint arXiv:2607.26244},
  year={2026},
  url={https://arxiv.org/abs/2607.26244}
}

@inproceedings{autocover,
  title={Automated Software Test Generation at Industry Scale Using a Multi-Agent Architecture and Workflow Integration},
  author={Rastenis, Matas and Chou, Ben and Choudhary, Shauvik Roy and Just, Ren{\'e}},
  booktitle={Proceedings of the IEEE/ACM 48th International Conference on Software Engineering: Software Engineering in Practice},
  pages={808--818},
  year={2026}
}

@article{llmsuite,
  title={Enhancing Automated Unit Test Generation for NLP Libraries Using Large Language Models},
  author={Deljouyi, Amirhossein and Panichella, Annibale and Zaidman, Andy},
  journal={arXiv preprint arXiv:2609.14784},
  year={2026}
}

@article{coveredcode,
  title={Enhancing llm-based test generation by eliminating covered code},
  author={Xu, WeiZhe and Liu, Mengyu and Kong, Fanxin},
  journal={arXiv preprint arXiv:2602.21997},
  year={2026}
}

@article{logichunter,
  title={LogicHunter: Testing LLM Agent Frameworks with an Agentic Oracle},
  author={Long, Minghui and Zhao, Yanjie and Wang, Haoyu},
  journal={Proceedings of the ACM on Software Engineering},
  volume={3},
  number={ISSTA},
  pages={1171--1193},
  year={2026},
  publisher={ACM New York, NY, USA}
}

@article{testdrivensurvey,
  title={Test-Driven Approaches to Software Engineering with Large Language Models: A Survey of Phases, Tasks, and Agent Skills},
  author={Liang, Yunhao and Gan, Chengguang and Ying, Ruixuan and Wei, Hanjun and Cui, Zhe and Ni, Shiwen},
  journal={arXiv preprint arXiv:2609.12012},
  year={2026}
}

@inproceedings{swebenchplus,
  title={SWE-Bench+: Enhanced LLM Coding Benchmark},
  author={Xue, Haoran and Aleithan, Reem and Enan, Nafid and Uddin, Gias and Wang, Song},
  booktitle={Proceedings of the 3rd ACM International Conference on AI-Powered Software},
  pages={332--339},
  year={2026}
}

@article{tebench,
  title={Breaking, Stale, or Missing? Benchmarking Coding Agents on Project-Level Test Evolution},
  author={Shang, Ye and Zhang, Quanjun and Hu, Haichuan and Fang, Chunrong and Xiao, Liang and Chen, Zhenyu},
  journal={arXiv preprint arXiv:2605.06125},
  year={2026}
}

@article{sweagent,
  title={Swe-agent: Agent-computer interfaces enable automated software engineering},
  author={Yang, John and Jimenez, Carlos and Wettig, Alexander and Lieret, Kilian and Yao, Shunyu and Narasimhan, Karthik and Press, Ofir},
  journal={Advances in Neural Information Processing Systems},
  volume={37},
  pages={50528--50652},
  year={2024}
}

@inproceedings{openhands,
  title={Openhands: An open platform for ai software developers as generalist agents},
  author={Wang, Xingyao and Li, Boxuan and Song, Yufan and Xu, Frank F and Tang, Xiangru and Zhuge, Mingchen and Pan, Jiayi and Song, Yueqi and Li, Bowen and Singh, Jaskirat and others},
  booktitle={International Conference on Learning Representations},
  volume={2025},
  pages={65882--65919},
  year={2025}
}

@inproceedings{swebench,
  title={Swe-bench: Can language models resolve real-world github issues?},
  author={Jimenez, Carlos E and Yang, John and Wettig, Alexander and Yao, Shunyu and Pei, Kexin and Press, Ofir and Narasimhan, Karthik},
  booktitle={International Conference on Learning Representations},
  volume={2024},
  pages={54107--54157},
  year={2024}
}

@article{swesmith,
  title={Swe-smith: Scaling data for software engineering agents},
  author={Yang, John and Lieret, Kilian and Jimenez, Carlos and Wettig, Alexander and Khandpur, Kabir and Zhang, Yanzhe and Hui, Binyuan and Press, Ofir and Schmidt, Ludwig and Yang, Diyi},
  journal={Advances in Neural Information Processing Systems},
  volume={38},
  year={2026}
}

@article{swecycle,
  title={SWE-Cycle: Benchmarking Code Agents across the Complete Issue Resolution Cycle},
  author={Guan, Hao and Fu, Lingyue and Zhang, Shao and Zhu, Yaoming and Zhang, Kangning and Qiu, Lin and Cai, Xunliang and Cao, Xuezhi and Liu, Weiwen and Zhang, Weinan and others},
  journal={arXiv preprint arXiv:2605.13139},
  year={2026}
}

@article{featurebench,
  title={Featurebench: Benchmarking agentic coding for complex feature development},
  author={Zhou, Qixing and Zhang, Jiacheng and Wang, Haiyang and Hao, Rui and Wang, Jiahe and Han, Minghao and Yang, Yuxue and Wu, Shuzhe and Pan, Feiyang and Fan, Lue and others},
  journal={arXiv preprint arXiv:2602.10975},
  year={2026}
}

@article{testbench,
  title={{TestBench}: Evaluating Class-Level Test Case Generation Capability of Large Language Models},
  author={Zhang, Quanjun and Shang, Ye and Fang, Chunrong and Gu, Siqi and Zhou, Jianyi and Chen, Zhenyu},
  journal={arXiv preprint arXiv:2409.17561},
  year={2024},
  url={https://arxiv.org/abs/2409.17561}
}

@article{clover,
  title={{CLOVER}: A Test Case Generation Benchmark with Coverage, Long-Context, and Verification},
  author={Xu, Jiacheng and Pang, Bo and Qu, Jin and Hayashi, Hiroaki and Xiong, Caiming and Zhou, Yingbo},
  journal={arXiv preprint arXiv:2502.08806},
  year={2025},
  url={https://arxiv.org/abs/2502.08806}
}

@inproceedings{testeval,
  title={Testeval: Benchmarking large language models for test case generation},
  author={Wang, Wenhan and Yang, Chenyuan and Wang, Zhijie and Huang, Yuheng and Chu, Zhaoyang and Song, Da and Zhang, Lingming and Chen, An Ran and Ma, Lei},
  booktitle={Findings of the Association for Computational Linguistics: NAACL 2025},
  pages={3547--3562},
  year={2025}
}

@inproceedings{testgeneval,
  title={Testgeneval: A real world unit test generation and test completion benchmark},
  author={Jain, Kush and Synnaeve, Gabriel and Rozi{\`e}re, Baptiste},
  booktitle={International Conference on Learning Representations},
  volume={2025},
  pages={14947--14999},
  year={2025}
}

@article{swtbench,
  title={Swt-bench: Testing and validating real-world bug-fixes with code agents},
  author={M{\"u}ndler, Niels and M{\"u}ller, Mark N and He, Jingxuan and Vechev, Martin},
  journal={Advances in Neural Information Processing Systems},
  volume={37},
  pages={81857--81887},
  year={2024}
}

@article{tddbench,
  title={Tdd-bench verified: Can llms generate tests for issues before they get resolved?},
  author={Ahmed, Toufique and Hirzel, Martin and Pan, Rangeet and Shinnar, Avraham and Sinha, Saurabh},
  journal={arXiv preprint arXiv:2412.02883},
  year={2024}
}

@article{reproagent,
  title={ReProAgent: Tool-Augmented Multi-Stage Agentic Generation of Bug Reproduction Tests from Issue Reports},
  author={Zhang, Quanjun and Zheng, Yi and Shang, Ye and Sun, Weifeng and Hu, Haichuan and Fang, Chunrong and Chen, Zhenyu and Xiao, Liang},
  journal={arXiv preprint arXiv:2607.09123},
  year={2026}
}

@article{swedoctor,
  title={SWE-Doctor: Guiding Software Engineering Agents with Runtime Diagnosis from Multi-Faceted Bug Reproduction Tests},
  author={Guo, Yaoqi and Liu, Yang and Zhang, Jie M and Ma, Yun and Lou, Yiling and Chen, Zhenpeng},
  journal={arXiv preprint arXiv:2607.00990},
  year={2026}
}

@article{testexplora,
  title={TestExplora: Benchmarking LLMs for Proactive Bug Discovery via Repository-Level Test Generation},
  author={Liu, Steven and Luo, Jane and Zhang, Xin and Liu, Aofan and Liu, Hao and Wu, Jie and Huang, Ziyang and Huang, Yangyu and Kang, Yu and Li, Scarlett},
  journal={arXiv preprint arXiv:2602.10471},
  year={2026}
}

@article{activeswe,
  title={Active-SWE: Benchmarking Coding Agents for Proactive Bug Fixing without Issue Reports},
  author={Li, Haobin and Deng, Ping and Qian, Weizhong and Jiang, Liang and Huang, Zhenyu and Yang, Mouxing and Peng, Xi},
  journal={arXiv preprint arXiv:2608.04682},
  year={2026}
}

@misc{swesweep,
  title={{SWE-sweep}: Can Agents Autonomously Find and Fix Bugs?},
  author={Lieret, Kilian and Ma, Jeffrey Jian and Kindi, Rahul and Wei, Yuxiang and Ma, Jeremy and Sootla, Sten and Thakkar, Parth and Zhou, Chao Beyond and Yin, Pengcheng and Hou, Rui and Press, Ofir and Yang, John},
  year={2026},
  note={Preprint},
  url={https://swesweep.com/paper}
}

@article{covmutation,
  title={Do Coverage and Mutation Scores of LLM-Generated Test Suites Correlate with Their Effectiveness?(Replicability Study)},
  author={Zhao, Junda and Zhou, Shurui and Cohen, Eldan},
  journal={arXiv preprint arXiv:2607.22880},
  year={2026}
}

@inproceedings{testoverfitting,
  title={Investigating Test Overfitting on SWE-bench},
  author={Ahmed, Toufique and Ganhotra, Jatin and Shinnar, Avraham and Hirzel, Martin},
  booktitle={Proceedings of the 34th ACM International Conference on the Foundations of Software Engineering},
  pages={1292--1296},
  year={2026}
}

\clearpage
\beginappendix

\section{Task Packaging and Instruction Specification}
\label{app:task_instructions}

\paragraph{The Harbor evaluation framework}
To ensure consistent task execution and reproducible evaluation, all tasks in \textbf{SpecEngineeringBench} are evaluated using the Harbor framework. Each task package contains the following components:
\begin{itemize}
  \setlength{\itemsep}{2pt}
  \item \path{task.toml}: Stores task metadata and execution configuration, with separate sections for the environment, agent, and verifier. These settings specify how Harbor runs the task, including applicable execution limits; they do not describe the defects the agent should discover.
  \item \path{instruction.md}: Defines the agent-facing objective, workspace paths, required actions, and submission format. In Task~1, it requests an executable test suite and test case metadata; in Task~2, it additionally requests execution evidence and a structured defect report. The task instructions are supplied alongside the instance-specific PRD, which describes the functionality to assess.
  \item \path{environment/}: Contains the Dockerfile and the repository and PRD inputs used to initialize the task container. The environment exposes the unrepaired repository at \path{/workspace/repo} and the requirements at \path{/workspace/prd.md}, with dependencies configured for test execution. Agent-authored tests and reports are written to \path{/workspace/submission/}.
  \item \path{tests/}: Contains the verifier entry point, \path{tests/test.sh}, and the scoring implementation for the corresponding task. Task~1 evaluates submitted tests on the defective and repaired snapshots to measure validity, defect detection, and coverage; Task~2 compares reported locations with reference annotations at file, function, and hunk granularities. These verifier resources are distinct from both the repository's existing test suite and the tests authored by the agent.
  \item \path{solution/}: Retains \path{solution/solve.sh}, the solution entry point in the Harbor package layout. It provides a separate location for a packaged solution script, while the evaluated agent writes its own test suite or report to the submission paths specified in \path{instruction.md}.
\end{itemize}
The agent works with the unrepaired repository; repaired snapshots, reference patches, and defect annotations are reserved for evaluation. Task~1 corresponds to the \path{task1/} package and Task~2 to the \path{task2/} package.

\paragraph{Task instruction specification.}
Each task specifies the agent's objective, workspace, and required submission. The PRD is available at \path{/workspace/prd.md}, the repository at \path{/workspace/repo}, and the submission directory at \path{/workspace/submission/}. The two boxes below reproduce the task prompts, including their submission requirements and JSON examples. Original Markdown syntax and indentation are retained; long lines wrap automatically for presentation. These are task-level instructions, distinct from the agent harness's system prompt; the instance-specific PRD is supplied separately.

\begin{promptbox}{Task 1: Test Generation}
You are an experienced software test engineer applying a Test-Driven Development (TDD) approach.
Your task is to design tests that ensure only correct implementations (following the documentation) pass, while
incorrect implementations would fail.

Your goal is to discover software problems by writing and executing test cases. Do not only describe tests or
return them in chat: save the executable tests in the task output directory, run them in the provided environment,
inspect the results, and revise the tests when necessary.

You are given the following information:
- Dependencies
- Test entry inferences
- Prd Documentation

Your tasks:
1. Infer the intended behavior of the repo from the prd documentation.
2. Design a set of comprehensive test cases that cover:
   - Basic functionality with valid inputs and expected outputs.
   - Boundary conditions and edge cases.
   - Invalid inputs and error handling.
   - Potential issues with dependency interactions.
3. Write executable test code using Pytest.
4. Ensure tests are designed to differentiate between correct and incorrect implementations:
   - At least one test should expose an incorrect implementation that does not fully follow the documented behavior.
   - A correct implementation should pass all tests.
5. Use all three complementary testing approaches to test the software comprehensively:
   - Unit testing for isolated function or method behavior.
   - Integration testing for interactions among repository components or dependencies.
   - Acceptance testing for externally observable end-to-end or user-facing behavior.
6. Save the complete executable test suite and any required test data under `/workspace/submission/`.
7. Use the environment instructions to configure the provided environment and run the complete test suite.
   Ensure the generated tests are executable with Pytest and that a correct implementation passes the suite.

## Exploration and completion protocol

Do not stop after writing a small representative sample of tests. Treat test design as an
iterative exploration task and work through the repository behavior systematically:

1. Build a checklist from every capability and public surface in the PRD. For each item,
   consider at least one normal case, one boundary or unusual case, and one invalid,
   failure, or state-transition case when that behavior is applicable. Include dependency
   interactions and externally observable workflows, not only isolated helper functions.
2. Work in multiple batches. For a non-trivial repository, plan several exploration
   batches; start each batch with some tests, then run the complete suite,
   inspect failures, and keep the passing tests. After the first batch, continue
   with the next unchecked PRD behavior instead of declaring success because the current
   tests pass.
3. Use coverage as feedback when it is available in the task environment. Compare the
   executed lines with the production entry functions and ranges described by the task
   information. For uncovered regions, decide whether they represent another behavior
   or a genuinely unreachable/non-behavioral line; add a focused test for the former and
   record the reason for the latter in your final notes.
4. Before stopping, perform a completion audit: every applicable PRD capability has a
   test attempt, all three categories have meaningful tests, the suite has been run from
   the task-local environment, and each test has a stable unique ID in
   `testcase_categories.json`. A passing suite alone is not a completion signal.

There is no fixed universal number of test cases: the required amount is determined by
the behavior checklist and the remaining coverage gaps. Continue until that checklist
and the coverage-driven review are complete, subject to the available execution budget.

## Environment

Use the preconfigured Harbor environment. Read the PRD at `/workspace/prd.md`, then run tests from the repository root: `/workspace/repo`


## Requirements for the generated tests
- Use Pytest and keep the complete suite executable in the provided repository environment; if the environment has problems, you may install the missing dependencies yourself.
- Apply unit, integration, and acceptance testing as complementary approaches for comprehensive coverage.
- Write `/workspace/submission/testcase_categories.json` with one object per testcase, containing `id` and exactly one
  `category` value from `unit_test`, `integration_test`, or `acceptance_test`. The JSON must cover every testcase
  in the executable suite and include all three testing approaches.
- Put test modules, fixtures, test data, and `testcase_categories.json` in `/workspace/submission/`.
- Run the complete suite after writing it by using the preconfigured environment described above.
- Do not modify production code, dependency manifests, the test runner configuration, or files outside the output
  directory.

- Write /workspace/submission/test_cases.json with id, category, and an exact pytest_nodeid for every testcase. The logical ID need not appear in the Python function name; parametrized cases must use their full collected node ID. Verify every node ID with pytest --collect-only.

## Output Format
The output is the complete executable test suite and `/workspace/submission/testcase_categories.json`.

\end{promptbox}

\begin{promptbox}{Task 2: Defect Discovery and Reporting}
You are an experienced software test engineer applying a systematic test-execution and bug-localization approach.
Your task is to inspect one repository and its neutral PRD, write and run focused temporary test cases against the provided repository,
identify source-grounded software problems, and produce a report.

You are given:
- PRD document
- The visible repository source
- The repository's declared dependencies

Your tasks:
1. Read the PRD and inspect the repository source, public interfaces, dependencies, and relevant documentation.
2. Design a test suite, execute it, analyze the run results, and try to uncover all hidden defects in the repository. It may cover:
   - Basic functionality with valid inputs and expected outputs.
   - Boundary conditions and edge cases.
   - Invalid inputs and error handling.
   - Potential issues with dependency interactions.
3. Write executable test code using Pytest. Base assertions on intended behavior so that correct implementations pass and incorrect implementations fail.
4. If there are issues with the runtime environment, you may perform fixes yourself, such as installing dependencies via pip install.
5. Save the complete executable test suite and any required test data under `/workspace/submission/`.
6. Use the environment instructions to configure the provided environment and run the complete test suite.
   Analyze observed results against expected behavior, investigate failures, and reproduce suspected bugs.
7. Trace confirmed failures to their causal source files, functions, and lines. Write a complete report
   to `/workspace/submission/report.json`, including test results, reproduction evidence, and root-cause analysis.

## Environment

The Harbor image exposes the model inputs at fixed paths. Read the PRD first from `/workspace/prd.md`, then inspect
and run the repository from `/workspace/repo`:

    cd /workspace/repo

Use the dependencies configured by `environment/Dockerfile`; do not recreate or replace the task environment (you can still fix the environment via pip install), and do not look for the repository or PRD beside the task directory. If setup has an infrastructure problem, record it in `constraints_and_disclosures` and do not claim it is a product failure without an independent reproduction.

## Output Format

The output is the complete executable test suite, execution evidence, and `/workspace/submission/report.json`.
When a targeted reproduction script is useful, include `/workspace/submission/repro.sh`.
The report's top-level structure must match this format:

```json
{
  "title": "short problem title",
  "status": "confirmed|suspected|blocked",
  "severity": "critical|high|medium|low|unknown",
  "test_results": [
    {
      "scope": "test scope",
      "cwd": "working directory",
      "command": "exact command",
      "exit_code": 0,
      "outcome": "what happened",
      "observed": "relevant output or failure",
      "evidence": "relative raw evidence path"
    }
  ],
  "reproduction": {
    "script": "repro.sh",
    "command": "/workspace/submission/repro.sh",
    "uses_authored_tests": true,
    "steps": ["reproduction step"],
    "expected": "documented or correct behavior",
    "actual": "observed incorrect behavior",
    "evidence": "relative raw evidence path",
    "environment_note": "environment limitation or intervention"
  },
  "locations": [
    {
      "path": "repository-relative/source.py",
      "functions": [
        {"name": "Class.method"}
      ],
      "lines": [42, 43],
      "finding": "source-grounded causal finding"
    }
  ],
  "root_cause": {
    "summary": "causal explanation",
    "mechanism": ["causal step"],
    "required_guard": "required behavioral or code guard, when applicable",
    "evidence": ["relative evidence path"]
  },
  "impact": "user-visible or system impact",
  "constraints_and_disclosures": ["setup, dependency, intervention, or scope disclosure"]
}
```

Use one JSON report object to describe the bugs you identify. Preserve exact commands, exit codes, outcomes, observed failures, and evidence paths. If an authored test run is blocked by the environment, record it as infrastructure evidence and do not claim it is a product failure without an independent reproduction.

## Localization Detail Requirements

The `locations` array must report every finding at all three localization granularities:

- File: `path` is the repository-relative source file.
- Function: include a `functions` array under that file, with each function's exact `name`. Use a qualified name such as `Class.method` when applicable; use `<module>` for module-level code.
- Code line: Each file entry must include a `lines` array containing the 1-based source-code line numbers of the exact lines relevant to the bug. Do not use vague ranges when precise line numbers can be identified.

The `path`, `functions`, and `lines` fields must be siblings within each file entry. The `lines` array describes relevant locations throughout that file, including locations outside function bodies. Do not nest `lines` inside individual function entries.

Use this shape (and repeat it for every affected file):

```json
"locations": [
  {
    "path": "pkg/module.py",
    "functions": [
      {"name": "<module>"},
      {"name": "Class.method"}
    ],
    "lines": [5, 42, 43]
  }
]
```

Always provide the file, function, and exact line details.

Do not produce a separate failure analysis or prose report in place of submission/report.json.
\end{promptbox}

\section{Foundation Models and Agent Harness Configurations}
\label{app:agent_configurations}

\paragraph{Models and harnesses.}
Our evaluation setup includes seven foundation models from five model families, paired with the coding-agent harnesses described in Section~\ref{sec:setup}. Each evaluated system combines a model with a harness that exposes repository inspection, file editing, and command execution, supporting an iterative cycle of test design, execution, and failure analysis. We use the following model--harness configurations:
\begin{itemize}
  \setlength{\itemsep}{2pt}
  \item \textbf{Anthropic Claude:} We evaluate Claude-Opus-5~\citep{claudeopus5} and Claude-Sonnet-5~\citep{claudesonnet5} through the Claude Code~\citep{claudecode}. Both models use its repository inspection, editing, and shell tools to author executable tests and investigate observed failures, providing two Claude configurations within the same harness family.
  \item \textbf{OpenAI GPT:} We evaluate GPT-6-Astra~\citep{gpt6astra} and GPT-5.6-Sol~\citep{gpt56sol} through the Codex~\citep{codexcli}. The models interact with the task workspace through Codex's file and command-execution tools, using test outcomes to guide subsequent exploration and report preparation.
  \item \textbf{Moonshot Kimi:} We evaluate Kimi-K3~\citep{kimik3} through Claude Code~\citep{claudecode} as the execution harness. This configuration extends the evaluation to the Kimi family while retaining the Claude Code~\citep{claudecode} tool interface used for the Claude configurations.
  \item \textbf{DeepSeek:} We evaluate DeepSeek-V4.1-Flash~\citep{deepseekv41flash} through Claude Code~\citep{claudecode}, using its repository inspection, editing, and shell tools under the same task-level input and submission contracts.
  \item \textbf{Z.ai GLM:} We evaluate GLM-5.3~\citep{glm53} through Claude Code~\citep{claudecode}, using the same task-level input and submission contracts as the other model families. This configuration extends the comparison of requirements-driven test generation and defect localization to the GLM family.
\end{itemize}

\paragraph{Task execution and submissions.}
Task~1 and Task~2 are evaluated independently: Task~2 does not receive the tests produced in Task~1. Both tasks use the instance's repository, PRD, and configured Docker environment. Task~1 submits executable tests and their registration metadata for evaluation on the defective and repaired snapshots. Task~2 submits independently authored tests, execution evidence, and a structured defect report. 


\clearpage
\section{Case Study}
\label{app:complete_case_study}

\textbf{Instance:} \path{python-visualization_folium/episode-0001} (five reference bugs).
\textbf{Model:} GPT-6-Astra with Codex.
The PRD below is the task input. Task~1 and Task~2 outputs come from independent runs on this same instance. We reproduce selected output excerpts; \texttt{...} marks omitted content, and test outcomes are supplied by the evaluator.


\subsection{Input: Product Requirements Document}

\begin{promptbox}[listing options={style=trajectoryexcerpt,breakatwhitespace=true},fonttitle=\sffamily\bfseries\normalsize,title after break={}]{PRD}
# Product Requirements Document

## Purpose

Folium is a Python library for creating interactive Leaflet.js maps. This PRD covers reliable JavaScript serialization, heatmap data handling, interactive drawing exports, TopoJSON overlays, and configurable Leaflet path styling for Python developers and integrators.

## User Interface And Interaction

Users create `folium.Map` objects and add layers or plugins through the public Python API, then render or save the resulting map as HTML. JavaScript options supplied through Folium configuration are emitted into the generated map. Users can add `folium.plugins.HeatMap` for weighted geographic points, `folium.plugins.Draw` for drawing and exporting GeoJSON, `folium.TopoJson` for TopoJSON overlays, and vector layers such as `PolyLine`, `Polygon`, `Circle`, and `CircleMarker` with Leaflet path options. The generated HTML exposes the corresponding Leaflet controls, overlays, styles, and downloads in a browser.

## Functional Requirements

### JavaScript serialization of dictionary options

- User goal: Configure map and layer options with Python dictionaries, including numeric keys.
- Primary interaction or public entry point: Pass dictionaries containing string or numeric keys to Folium options rendered with the `tojavascript` template filter.
- Expected outcome: The generated JavaScript is valid and preserves every dictionary key and value, including numeric keys, so Leaflet receives the intended options.
- User-visible errors or constraints: Values must remain JSON/JavaScript-serializable according to Folium's supported option types.
- Externally visible state change, when applicable: Rendered HTML contains the complete option object used by the browser.

### Weighted HeatMap data

- User goal: Display geographic points with optional intensity weights from Python lists, pandas-compatible data, or NumPy arrays.
- Primary interaction or public entry point: Instantiate `folium.plugins.HeatMap(data)` with rows shaped as `[latitude, longitude]` or `[latitude, longitude, weight]`.
- Expected outcome: Valid integer or floating-point NumPy weights are accepted and rendered as heatmap intensities without changing their geographic locations.
- User-visible errors or constraints: Coordinates must pass Folium location validation; data containing NaN values is rejected with a clear value error. The input must provide at least latitude and longitude for each point.
- Externally visible state change, when applicable: The generated map contains a Leaflet heat layer whose points and intensities reflect the supplied data.

### Draw export controls

- User goal: Draw and edit geometries on a map and download the collected shapes as GeoJSON.
- Primary interaction or public entry point: Add `folium.plugins.Draw(export=True)` to a map, optionally configuring `filename`, control `position`, `export_position`, drawing options, editing options, and event handlers.
- Expected outcome: Each enabled Draw instance displays a usable export link with a unique DOM id. Activating it prepares a download containing the current drawn feature group as GeoJSON, using the configured filename.
- User-visible errors or constraints: `export_position` accepts only `topright`, `topleft`, `bottomright`, or `bottomleft`; unsupported values produce a value error. Export is available only when enabled.
- Externally visible state change, when applicable: Drawn layers are added to the editable feature group, and the uniquely identified export link receives a data download target when activated. Multiple Draw controls on one map operate independently.

### TopoJSON geometry overlays

- User goal: Render a selected TopoJSON object on a Folium map, regardless of whether the object is represented as a collection or a standalone geometry.
- Primary interaction or public entry point: Instantiate `folium.TopoJson(data, object_path, ...)` with a mapping, an embedded file-like source, or another documented string source, and identify the desired object path.
- Expected outcome: The selected TopoJSON object is converted to a Leaflet GeoJSON layer and displayed, with optional styling, tooltips, layer-control settings, and smoothing.
- User-visible errors or constraints: The supplied object path and TopoJSON data must identify a supported TopoJSON object; invalid data or paths are reported during normal data loading or rendering.
- Externally visible state change, when applicable: A visible map overlay is created and can participate in layer controls, styling, and tooltips.

### Leaflet path option forwarding

- User goal: Use the full set of Leaflet path configuration options when creating Folium vector overlays.
- Primary interaction or public entry point: Pass documented snake_case or lowerCamelCase path options, including additional Leaflet options, through vector-layer constructors or `path_options`.
- Expected outcome: Recognised options are normalized to Leaflet names and unrecognised but valid Leaflet path options are retained and emitted to the browser instead of being discarded.
- User-visible errors or constraints: Option values must be compatible with Leaflet and JSON/JavaScript serialization. Explicit fill and color settings follow the documented precedence of the public API.
- Externally visible state change, when applicable: The rendered vector layer reflects all supplied path behavior, appearance, interaction, and placement options.

## Non-Functional Requirements

- Compatibility: Support the package's documented Python requirement of version 3.9 or newer and remain compatible with the Leaflet.js and plugin integrations used by Folium's generated HTML.
- Data fidelity: Preserve numeric values and option keys through Python-to-JavaScript rendering without silently dropping supported input.
- Validation and stability: Reject invalid geographic coordinates and NaN heatmap data predictably, and report invalid Draw control placement values clearly.
- Interoperability: Accept the documented data forms for heatmaps and TopoJSON, including NumPy-compatible point data, mappings, and file-like inputs.
- Isolation: Multiple interactive controls or overlays placed on the same map must retain independent browser identifiers and behavior.
- Resource and delivery model: Generated maps must remain self-contained with respect to embedded user data while loading the documented Leaflet and plugin resources required for browser interaction.
\end{promptbox}

\clearpage
\subsection{Task 1: Generated Tests}

The submission contains 159 tests (85 unit, 60 integration, 14 acceptance). Evaluator results: HP 83.65\%, F2P 20.75\%, EC 96.30\%, and CFG 100.00\%. There are 33 F2P tests, 26 tests that fail on the repaired snapshot, and 100 remaining tests that do not expose a reference defect (including two skipped DataFrame cases). The excerpts below retain the submitted function bodies; parameter lists, imports, helpers, and other tests are omitted.

\begin{trajectorycase}[breakable=false,before upper={\raggedright\setlength{\parskip}{4pt}}]{casediscovery}{Generated tests that expose reference defects}
\textbf{Numeric dictionary keys} (\path{test_serialization_unit.py}; selected parameter: \texttt{key=2}).
\begin{lstlisting}[style=trajectoryexcerpt]
def test_numeric_dictionary_keys_are_preserved(key):
    assert run_js(tojavascript({key: 'color', 'other': 4})) == {str(key): 'color', 'other': 4}
\end{lstlisting}

\textbf{Defective:} fail, \texttt{AttributeError: 'int' object has no attribute 'split'}.
\textbf{Repaired:} pass. \textbf{F2P:} yes.

\textbf{Additional Leaflet options} (\path{test_path_unit.py}; selected parameters: \texttt{key='pane'}, \texttt{value='customPane'}, \texttt{normalized='pane'}).
\begin{lstlisting}[style=trajectoryexcerpt]
def test_additional_leaflet_path_options_survive(key, value, normalized):
    assert path_options(**{key: value})[normalized] == value
\end{lstlisting}

\textbf{Defective:} fail, \texttt{KeyError: 'pane'}.
\textbf{Repaired:} pass. \textbf{F2P:} yes.
\begin{lstlisting}[style=trajectoryexcerpt]
...
\end{lstlisting}
\end{trajectorycase}

\begin{trajectorycase}[breakable=false,before upper={\raggedright\setlength{\parskip}{4pt}}]{casegap}{Generated tests that also fail on the repaired snapshot}
\textbf{Quoted dictionary keys} (\path{test_serialization_unit.py}; selected parameter: \texttt{key='a"b'}).
\begin{lstlisting}[style=trajectoryexcerpt]
def test_string_keys_produce_valid_javascript(key):
    assert run_js(tojavascript({key: 3})) == {key: 3}
\end{lstlisting}

\textbf{Defective / repaired:} fail / fail, \texttt{SyntaxError: Unexpected identifier 'b'}.
\textbf{F2P:} no.

\textbf{Draw export placement} (\path{test_draw_unit.py}; selected parameter: \texttt{position='topright'}).
\begin{lstlisting}[style=trajectoryexcerpt]
def test_all_documented_export_positions_are_accepted(position):
    draw = Draw(export=True, export_position=position)
    assert draw.export_position == position
\end{lstlisting}

\textbf{Defective / repaired:} fail / fail, \texttt{TypeError: Draw.\_\_init\_\_() got an unexpected keyword argument 'export\_position'}.
\textbf{F2P:} no.
\begin{lstlisting}[style=trajectoryexcerpt]
...
\end{lstlisting}
\end{trajectorycase}

\begin{trajectorycase}[breakable=false,before upper={\raggedright\setlength{\parskip}{4pt}}]{caseboundary}{Generated tests that pass on both snapshots}
\textbf{Ordinary JSON values} (\path{test_serialization_unit.py}; selected parameter: \texttt{value=0}).
\begin{lstlisting}[style=trajectoryexcerpt]
def test_json_values_round_trip(value):
    assert run_js(tojavascript(value)) == value
\end{lstlisting}

\textbf{Defective / repaired:} pass / pass. \textbf{F2P:} no.

\textbf{Boolean options} (\path{test_path_unit.py}; selected parameter: \texttt{gradient=False}).
\begin{lstlisting}[style=trajectoryexcerpt]
def test_gradient_boolean_preserved(gradient):
    assert path_options(gradient=gradient)['gradient'] is gradient
\end{lstlisting}

\textbf{Defective / repaired:} pass / pass. \textbf{F2P:} no.
\begin{lstlisting}[style=trajectoryexcerpt]
...
\end{lstlisting}
\end{trajectorycase}

\clearpage
\subsection{Task 2: Defect Report}

The independently authored Task~2 suite has 64 failed tests, 29 passed tests, and seven pandas-only skips. The submitted report contains 15 agent-reported findings. Evaluator results: five of five reference bugs localized; file/function/hunk LR = 100.00\%/100.00\%/100.00\%, and LP = 83.33\%/40.00\%/62.50\%. The excerpt retains the report's wording, with omitted fields, findings, and explanations marked by \texttt{...}.

\begin{promptbox}[listing options={style=trajectoryexcerpt,breakatwhitespace=true},fonttitle=\sffamily\bfseries\normalsize,title after break={}]{Task 2 output: report.json}
{
  "title": "15 confirmed Folium serialization, heatmap, Draw, TopoJSON, and path-option defects",
  "status": "confirmed",
  "severity": "high",
  "test_results": [
    {
      "scope": "Final complete authored Pytest suite: 100 collected cases across all five PRD areas",
      "cwd": "/workspace/repo",
      "command": "/workspace/submission/repro.sh > /workspace/submission/evidence/run-console.log 2>&1",
      "exit_code": 1,
      "outcome": "64 failed, 29 passed, 7 skipped; no collection/runtime infrastructure errors in the final suite. Failures represent 15 source-grounded findings, not 64 separate bugs.",
      "observed": "Draw: 10 failed/3 passed. HeatMap: 13 failed/14 passed/7 optional pandas skips. Serialization: 10 failed/4 passed. TopoJSON: 22 failed/5 passed. Vectors: 9 failed/3 passed. The script executes: python -m pytest /workspace/submission -vv --tb=short -rA --junitxml=/workspace/submission/evidence/results.xml .",
      "evidence": "evidence/pytest.log"
    },
    ...
  ],
  "reproduction": {
    "script": "repro.sh",
    "command": "/workspace/submission/repro.sh",
    "uses_authored_tests": true,
    ...
    "expected": "All supported serialization, heatmap, drawing, TopoJSON, and vector-option cases pass; deliberately invalid inputs produce the required errors. Independent Draw links preserve feature-group and filename identity.",
    "actual": "64 assertion/exception failures confirm B01-B15. Normal string options, supported JsCode/Element references, ordinary heatmaps and location/NaN validation, single Draw export/event configuration, collection TopoJSON styling/tooltips, and documented vector normalization/fill precedence pass. Seven DataFrame cases are skipped solely for missing pandas.",
    "evidence": "evidence/pytest.log",
    ...
  },
  "locations": [
    {"path": "folium/template.py",
     "functions": [{"name":"tojavascript"}],
     "lines": [17,18],
     "finding": "B01 [high]: Numeric dictionary keys crash JavaScript option serialization. ..."},
    {"path": "folium/plugins/heat_map.py",
     "functions": [{"name":"HeatMap.__init__"},{"name":"<module>"}],
     "lines": [51,80,81,82],
     "finding": "B03 [high]: NumPy heatmap intensities are not made JSON serializable. ..."},
    {"path": "folium/plugins/draw.py",
     "functions": [{"name":"<module>"}],
     "lines": [61,77,126,127,130,133,134],
     "finding": "B05 [high]: Multiple Draw exports share one DOM id and overwrite each other. ..."},
    {"path": "folium/features.py",
     "functions": [{"name":"TopoJson.style_data"},{"name":"TopoJson.render"}],
     "lines": [1056,1057,1059,1066],
     "finding": "B07 [high]: TopoJson rendering unconditionally assumes a GeometryCollection. ..."},
    {"path": "folium/features.py",
     "functions": [{"name":"GeoJsonDetail.render"}],
     "lines": [1202,1203,1204,1205,1206,1207],
     "finding": "B13 [medium]: TopoJSON tooltips independently assume a nonempty geometry collection. ..."},
    {"path": "folium/vector_layers.py",
     "functions": [{"name":"path_options"}],
     "lines": [81,110,123,125,126],
     "finding": "B14 [medium]: Valid additional Leaflet path options are silently discarded. ..."},
    ...
  ],
  "root_cause": {
    "summary": "The confirmed failures arise from unchecked input-shape/type assumptions, incomplete option forwarding, and non-isolated generated identifiers. The 15 findings are separated below by their causal source logic; repeated parameterized failures are counted only once per finding.",
    ...
  },
  "impact": "Valid maps fail during Python rendering for numeric option keys, common NumPy heatmap weights, and supported TopoJSON forms. Escaping defects can corrupt property names or break generated scripts. Multiple Draw exports can download the wrong feature group and filename, while one link is unusable. Export placement cannot be configured. Additional vector options silently lose interaction, placement, and attribution settings; show=False overlays remain initially visible; TopoJSON bounds can fail or encompass unrelated data. Severity is high overall for rendering and export/data-fidelity failures; individual severities are recorded in the finding entries.",
  "constraints_and_disclosures": [
    "Pandas, declared in requirements-dev.txt, was unavailable. Both installation attempts failed with package-index TLS/network errors. The initial collection error is preserved, and seven DataFrame-only tests skip explicitly. No pandas failure is claimed; NumPy/list reproductions independently confirm B03.",
    "Draw handlers and TopoJSON visibility were evaluated in Node.js against documented DOM/Leaflet interfaces implemented by deterministic test doubles. These establish generated-code behavior and duplicate-id effects, but are not full-browser visual, real-download, or CDN integration tests.",
    ...
  ]
}
\end{promptbox}


\clearpage
\section{Verbatim Trajectory Case Studies and Failure Mechanics}
\label{app:trajectory_case_studies}

Complementing the complete walkthrough in Appendix~\ref{app:complete_case_study}, the following six cases provide focused qualitative evidence for the mechanisms discussed in RQ1 and RQ5. Cases~1--2 form a paired GPT-5.6-Sol comparison on the same Folium instance: the repository-provided suite leaves an option contract unchecked, whereas requirements-derived assertions expose the shared defect. Case~3 adds a Textual pair in which thousands of passing existing tests coexist with boundary failures exposed by authored tests. Cases~4--6 examine GPT-6-Astra test-generation outcomes. Each case combines task metadata, operational outcome, verbatim commands or test excerpts, and a short failure or success analysis. Excerpts are transcribed from saved trajectories, submitted tests, and execution logs; step numbers refer to the original trajectories. 


\phantomsection\label{casegap}
\begin{trajectorycase}[breakable=false]{casegap}{\faTimesCircle\hspace{0.4em}Case 1: Passing Tests Leave an Option Contract Unchecked (GPT-5.6-Sol on Folium)}
\textbf{Task Metadata:} \path{python-visualization_folium/episode-0004} (Python; archived evaluation).\\
\textbf{Agent \& Harness:} GPT-5.6-Sol via Codex.\\
\textbf{Test Suite:} Repository-provided tests.\\
\textbf{Execution Evidence:} PRD-focused subset: 30 passed, zero failed.\\
\textbf{Localization Recall (File / Function / Hunk):} 20.00\% / 0.00\% / 0.00\%.\\
\textbf{Outcome:} The report omits the recorded option-forwarding defect.

\textbf{Verbatim Test Command and Output (Step 25).}
The agent runs the following command in \path{/workspace/repo}; the output is saved in \path{evidence/prd_suite.raw.txt} with exit code zero.
\begin{lstlisting}[style=trajectoryexcerpt]
python -m pytest -q tests/plugins/test_heat_map.py tests/test_features.py tests/test_vector_layers.py

..............................                                           [100%]
30 passed in 1.15s
\end{lstlisting}

\textbf{Verbatim Agent Narration (Step 28, excerpt).}
\begin{quote}
\itshape ``Focused reproduction matches both failures exactly, and the PRD-targeted test set passes 30/30.''
\end{quote}

\textbf{Failure Analysis:} The existing vector-layer tests check default configurations and conversion of \texttt{fill\_color} to \texttt{fillColor}, but contain no assertion that extra options such as \texttt{pane} are retained. Thus, passing tests of the relevant components leave this explicit requirement unchecked. The agent reports failures concerning stale map CSS and non-string icon colors, while the shared option-forwarding defect remains absent from its report.

\end{trajectorycase}

\phantomsection\label{casediscovery}
\begin{trajectorycase}[breakable=false]{casediscovery}{\faCheckCircle\hspace{0.4em}Case 2: Requirements-Derived Assertions Expose a Shared Defect (GPT-5.6-Sol on Folium)}
\textbf{Task Metadata:} \path{python-visualization_folium/episode-0004} (Python; archived evaluation).\\
\textbf{Agent \& Harness:} GPT-5.6-Sol via Codex.\\
\textbf{Test Suite:} Agent-authored tests.\\
\textbf{Execution Evidence:} Five extra-option cases fail; all five fail again on reproduction.\\
\textbf{Localization Recall (File / Function / Hunk):} 80.00\% / 71.43\% / 71.43\%.\\
\textbf{Outcome:} The report localizes the shared defect to \texttt{path\_options}.

\textbf{Verbatim Submitted Test (written at Step 13, excerpt).}
The agent hypothesizes at Step~8 that \texttt{path\_options} drops unconsumed keywords. It then parameterizes the assertion below over \texttt{PolyLine}, \texttt{Polygon}, \texttt{Rectangle}, \texttt{Circle}, and \texttt{CircleMarker}. The excerpt preserves lines 210--224 of \path{test_prd_behaviors.py}; the parameterization is omitted.
\begin{lstlisting}[style=trajectoryexcerpt]
def test_vector_layers_preserve_additional_leaflet_path_options(factory):
    layer = factory()

    assert layer.options["pane"] == "labels"
    assert layer.options["renderer"] == "canvas-renderer"
    assert layer.options["interactive"] is False
    assert layer.options["customOption"] == 7


def _extension_options():
    return {
        "pane": "labels",
        "renderer": "canvas-renderer",
        "interactive": False,
        "custom_option": 7,
    }
\end{lstlisting}

\textbf{Verbatim Test Output (Step 16, PolyLine case).}
The 27-case suite has six passes and 21 failures, with zero JUnit errors. All five extra-option cases fail at the same assertion; \path{full-run.txt} records:
\begin{lstlisting}[style=trajectoryexcerpt]
>       assert layer.options["pane"] == "labels"
               ^^^^^^^^^^^^^^^^^^^^^
E       KeyError: 'pane'

../submission/test_prd_behaviors.py:212: KeyError
\end{lstlisting}

\textbf{Success Analysis:} The new assertion directly checks the PRD's extra-option contract. The agent traces the failures to \path{folium/vector_layers.py}, function \texttt{path\_options}, including lines 123, 125, and 126. This helper consumes named options and merges \texttt{extra\_options}, but discards the remaining \texttt{kwargs}; all five constructors depend on it. The reference fix for \path{python-visualization_folium-pr-2272} adds \texttt{default.update(kwargs)} before the return, agreeing with this diagnosis. The five failures therefore exercise one shared recorded defect.

\end{trajectorycase}

\phantomsection
\label{app:rq4_textual}
\phantomsection\label{caseboundary}
\begin{trajectorycase}[breakable=false]{caseboundary}{\faInfoCircle\hspace{0.4em}Case 3: Thousands of Passing Tests Miss Boundary Failures (GPT-5.6-Sol on Textual)}
\textbf{Task Metadata:} \path{textual/episode-0001} (Python).\\
\textbf{Agent \& Harness:} GPT-5.6-Sol via Codex, paired testing conditions.\\
\textbf{Execution Evidence:} Existing suite: 3,448 passed; authored suite: 19 passed, three failed.\\
\textbf{Localization Recall (File / Function / Hunk):} Repository-provided: 0.00\% / 0.00\% / 0.00\%; agent-authored: 18.18\% / 9.52\% / 2.78\%.\\
\textbf{Outcome:} Authored boundary checks yield failures and nonzero reference agreement.

\textbf{Verbatim Existing-Suite Command and Output.}
The initial run has 437 failures under \texttt{NO\_COLOR=1} and \texttt{TERM=dumb}. The agent adjusts only these terminal-color variables and reruns the full suite. The command and final output in \path{evidence/full_pytest_color_enabled_output.txt} are:
\begin{lstlisting}[style=trajectoryexcerpt]
env -u NO_COLOR TERM=xterm-256color python -m pytest -q

445 snapshots passed.
3448 passed, 3 skipped, 4 xfailed, 1 xpassed in 292.80s (0:04:52)
\end{lstlisting}

\textbf{Verbatim Authored Test (submitted source, lines 189--195).}
The same PRD requires Rich-compatible rendering of plain and styled text. The new suite includes this boundary check for an unstyled renderable at zero opacity:
\begin{lstlisting}[style=trajectoryexcerpt]
def test_zero_text_opacity_handles_unstyled_rich_segments() -> None:
    console = Console(width=20, color_system="truecolor")
    renderable = TextOpacity(Text("plain", end=""), opacity=0)

    rendered = list(console.render(renderable))

    assert "".join(segment.text for segment in rendered) == " " * len("plain")
\end{lstlisting}

\textbf{Verbatim Failure Output (\path{pytest-full.txt}, excerpt).}
\begin{lstlisting}[style=trajectoryexcerpt]
>               invisible_style = _from_color(bgcolor=style.bgcolor)
                                                      ^^^^^^^^^^^^^
E               AttributeError: 'NoneType' object has no attribute 'bgcolor'

src/textual/renderables/text_opacity.py:80: AttributeError

3 failed, 19 passed in 1.07s
\end{lstlisting}

\textbf{Analysis:} The zero-opacity branch assumes every Rich segment has a style, although valid unstyled text produces \texttt{style=None}. The other two failures concern an out-of-range CSI-u Unicode value and an rxvt control-down sequence decoded as control-up. Focused reproductions repeat the rendering failure and both parser failures; a separate run has 16 passing controls. The agent reports the parser table, parser function, and rendering branch. The pair shows that a large passing suite can leave specific requirements-derived boundary checks unresolved, while the low hunk recall also shows that these failures do not yield comprehensive reference localization.

\end{trajectorycase}

\phantomsection\label{caseastrweak}
\begin{trajectorycase}[breakable=false]{caseastrweak}{\faTimesCircle\hspace{0.4em}Case 4: Broad Test Registration Without Defect Exposure (GPT-6-Astra on Apache Superset)}
\textbf{Task Metadata:} \path{apache_superset/episode-0003} (Python).\\
\textbf{Agent \& Harness:} GPT-6-Astra via Codex.\\
\textbf{Operational Metrics:} HP 100.00\% / F2P 0.00\% / EC 56.82\% / CFG 69.23\% / Num 99.\\
\textbf{Outcome:} The registered suite is compatible with the repaired snapshot but exposes none of the recorded defects.

\textbf{Verbatim Agent Narration (Step 40).}
\begin{quote}
\itshape ``The suite now runs end to end: 99 tests collect, with 17 passing and 64 expected failures that directly expose supplied implementation regressions; 18 xpasses show behaviors already fixed in this checkout.''
\end{quote}

\textbf{Failure Analysis:} The agent invests in broad unit, integration, and acceptance coverage and reports a clean collection, yet the verifier records no F2P test. The trajectory treats expected-failure outcomes and xpasses as evidence of progress, but those outcomes do not establish a test that separates the defective and repaired snapshots. This case isolates the RQ1 gap between a large, executable submission and defect-discriminating assertions.

\end{trajectorycase}

\phantomsection\label{caseastrasuccess}
\begin{trajectorycase}[breakable=false]{caseastrasuccess}{\faCheckCircle\hspace{0.4em}Case 5: Requirements-Derived Assertions Expose Multiple Defects (GPT-6-Astra on Folium)}
\textbf{Task Metadata:} \path{python-visualization_folium/episode-0004} (Python; archived evaluation).\\
\textbf{Agent \& Harness:} GPT-6-Astra via Codex.\\
\textbf{Operational Metrics:} HP 95.54\% / F2P 33.12\% / EC 88.57\% / CFG 100.00\% / Num 157.\\
\textbf{Outcome:} Requirements-derived tests expose multiple defective behaviors while retaining high validity and change coverage.

\textbf{Verbatim Agent Narration (Step 9).}
\begin{quote}
\itshape ``The code suggests several gaps worth testing: NumPy weights may remain non-serializable, TopoJSON styling assumes a geometry collection, and extra vector options appear to be dropped.''
\end{quote}

\textbf{Verbatim Submitted Assertion (Step 10, excerpt).}
\begin{lstlisting}[style=trajectoryexcerpt]
def test_each_vector_constructor_forwards_options(kind):
    layer = vector(kind, interactive=False, pane="customPane", ...)
    assert layer.options["interactive"] is False
\end{lstlisting}

\textbf{Success Analysis:} The agent converts the PRD into concrete serialization, geometry, and option-forwarding checks. On the defective snapshot, the focused run reports \texttt{KeyError: 'interactive'} for the vector-option assertion; the full submission records 99 passes and 58 failures, and the verifier credits 33.12\% of registered tests as F2P. The case shows how a requirements-derived oracle can turn broad code coverage into discriminating defect exposure.

\end{trajectorycase}

\phantomsection\label{caseastraassert}
\begin{trajectorycase}[breakable=false]{caseastraassert}{\faExclamationTriangle\hspace{0.4em}Case 6: Assertion Oracle Mismatch Under a Familiar API (GPT-6-Astra on Ax)}
\textbf{Task Metadata:} \path{facebook_Ax/episode-0004} (Python).\\
\textbf{Agent \& Harness:} GPT-6-Astra via Codex.\\
\textbf{Operational Metrics:} HP 19.26\% / F2P 3.70\% / EC 61.04\% / CFG 20.00\% / Num 135.\\
\textbf{Outcome:} Many tests fail on the repaired snapshot because their expected transformed values do not match the public API behavior.

\textbf{Verbatim Failing Assertion (\path{test_standardization.py}).}
\begin{lstlisting}[style=trajectoryexcerpt]
assert oc.objective.weights == pytest.approx(
    [weights[0] * 2, weights[1] * 20]
)
E assert [2.0, -3.0] == approx([4.0, -60.0])
\end{lstlisting}

\textbf{Failure Analysis:} The test encodes an inferred rescaling rule that conflicts with the documented contract. The same assertion passes on the defective snapshot but fails on the repaired snapshot, so it cannot count as F2P and instead lowers HP. The trajectory's final audit reports 135 tests, 91 passes, and 44 failures; the verifier credits only 3.70\% F2P. This case exposes assertion mismatch as a direct construction failure even when the suite exercises the target transformation code.

\end{trajectorycase}


\section{Repository Inventory and Instance Counts}
\label{app:task_catalog}
Table~\ref{tab:task_catalog} lists 78 repositories, 100 instances, and 531 recorded defect occurrences. Instances from the same upstream repository are grouped together. Each instance supports both Task~1 and Task~2 and is counted once.
\begingroup
\fontsize{8pt}{10pt}\selectfont
\setlength{\tabcolsep}{4pt}
\renewcommand{\arraystretch}{1.12}
\setlength{\LTleft}{\fill}
\setlength{\LTright}{\fill}
\begin{longtable}{@{}>{\raggedright\arraybackslash}p{0.48\textwidth}cccc@{}}
\caption{Repository inventory by application type. License lists the current upstream primary code license, checked on 15 October 2026 for retained repositories and 20 October 2026 for additions, with links to the upstream terms. Defects are recorded occurrences in the current reference annotations, counted separately in each instance. kLOC is the median non-documentation source-code line count across each repository's released defective snapshots, expressed in thousands of lines ($1\,\mathrm{kLOC}=1{,}000$ lines) and rounded to one decimal place. Counts exclude comments and blank lines while retaining tests, examples, generated code, and vendored source.}\label{tab:task_catalog}\\
\toprule
\textbf{Repository} & \textbf{License} & \textbf{Instances} & \textbf{Defects} & \textbf{kLOC} \\
\midrule
\endfirsthead
\multicolumn{5}{l}{\textit{Table~\ref{tab:task_catalog} continued}}\\
\toprule
\textbf{Repository} & \textbf{License} & \textbf{Instances} & \textbf{Defects} & \textbf{kLOC} \\
\midrule
\endhead
\midrule
\multicolumn{5}{r}{\textit{Continued on next page}}\\
\endfoot
\bottomrule
\endlastfoot
\rowcolor{dividergray}
\multicolumn{5}{c}{\textbf{\textit{Communications}}} \\*
\href{https://github.com/element-hq/synapse}{element-hq/\allowbreak synapse} & \href{https://github.com/element-hq/synapse/blob/develop/LICENSE-AGPL-3.0}{AGPL-3.0} & 1 & 13 & 283.0 \\*
\href{https://github.com/letta-ai/letta}{letta-ai/\allowbreak letta} & \href{https://github.com/letta-ai/letta/blob/main/LICENSE}{Apache-2.0} & 1 & 5 & 161.8 \\
\rowcolor{dividergray}
\multicolumn{5}{c}{\textbf{\textit{Data Analysis}}} \\*
\href{https://github.com/apache/superset}{apache/\allowbreak superset} & \href{https://github.com/apache/superset/blob/master/LICENSE.txt}{Apache-2.0} & 4 & 26 & 1,017.0 \\*
\href{https://github.com/bashtage/arch}{bashtage/\allowbreak arch} & \href{https://github.com/bashtage/arch/blob/main/LICENSE.md}{NCSA} & 2 & 10 & 42.2 \\*
\href{https://github.com/comet-ml/opik}{comet-ml/\allowbreak opik} & \href{https://github.com/comet-ml/opik/blob/main/LICENSE}{Apache-2.0} & 1 & 5 & 1,005.1 \\*
\href{https://github.com/Data-Centric-AI-Community/data-profiling}{Data-Centric-AI-Community/\allowbreak data-profiling} & \href{https://github.com/Data-Centric-AI-Community/fg-data-profiling/blob/develop/LICENSE}{MIT} & 1 & 4 & 16.5 \\*
\href{https://github.com/e2nIEE/pandapower}{e2nIEE/\allowbreak pandapower} & \href{https://github.com/e2nIEE/pandapower/blob/develop/LICENSE}{BSD-3-Clause} & 1 & 5 & 84.6 \\*
\href{https://github.com/lmfit/lmfit-py}{lmfit/\allowbreak lmfit-py} & \href{https://github.com/lmfit/lmfit-py/blob/master/LICENSE}{BSD-3-Clause} & 1 & 4 & 10.7 \\*
\href{https://github.com/python-visualization/folium}{python-visualization/\allowbreak folium} & \href{https://github.com/python-visualization/folium/blob/main/LICENSE.txt}{MIT} & 2 & 11 & 6.5 \\*
\href{https://github.com/pyvista/pyvista}{pyvista/\allowbreak pyvista} & \href{https://github.com/pyvista/pyvista/blob/main/LICENSE}{MIT} & 1 & 7 & 119.8 \\*
\href{https://github.com/skforecast/skforecast}{skforecast/\allowbreak skforecast} & \href{https://github.com/skforecast/skforecast/blob/main/LICENSE}{BSD-3-Clause} & 1 & 4 & 112.8 \\
\rowcolor{dividergray}
\multicolumn{5}{c}{\textbf{\textit{Database}}} \\*
\href{https://github.com/airbytehq/airbyte}{airbytehq/\allowbreak airbyte} & \href{https://github.com/airbytehq/airbyte/blob/master/LICENSE}{ELv2} & 1 & 4 & 525.5 \\*
\href{https://github.com/ansible/awx}{ansible/\allowbreak awx} & \href{https://github.com/ansible/awx/blob/devel/LICENSE.md}{Apache-2.0} & 1 & 5 & 113.2 \\*
\href{https://github.com/coleifer/peewee}{coleifer/\allowbreak peewee} & \href{https://github.com/coleifer/peewee/blob/master/LICENSE}{MIT} & 1 & 5 & 35.5 \\*
\href{https://github.com/dagster-io/dagster}{dagster-io/\allowbreak dagster} & \href{https://github.com/dagster-io/dagster/blob/master/LICENSE}{Apache-2.0} & 1 & 4 & 1,107.0 \\*
\href{https://github.com/datahub-project/datahub}{datahub-project/\allowbreak datahub} & \href{https://github.com/datahub-project/datahub/blob/master/LICENSE}{Apache-2.0} & 2 & 13 & 1,760.3 \\*
\href{https://github.com/dgunning/edgartools}{dgunning/\allowbreak edgartools} & \href{https://github.com/dgunning/edgartools/blob/main/LICENSE.txt}{MIT} & 3 & 14 & 922.5 \\*
\href{https://github.com/django-haystack/django-haystack}{django-haystack/\allowbreak django-haystack} & \href{https://github.com/django-haystack/django-haystack/blob/master/LICENSE}{BSD-3-Clause} & 1 & 5 & 15.5 \\*
\href{https://github.com/facebook/Ax}{facebook/\allowbreak Ax} & \href{https://github.com/facebook/Ax/blob/main/LICENSE}{MIT} & 3 & 14 & 150.7 \\*
\href{https://github.com/Farama-Foundation/Gymnasium}{Farama-Foundation/\allowbreak Gymnasium} & \href{https://github.com/Farama-Foundation/Gymnasium/blob/main/LICENSE}{MIT} & 1 & 6 & 32.0 \\*
\href{https://github.com/mealie-recipes/mealie}{mealie-recipes/\allowbreak mealie} & \href{https://github.com/mealie-recipes/mealie/blob/mealie-next/LICENSE}{AGPL-3.0} & 1 & 5 & 107.3 \\*
\href{https://github.com/mlflow/mlflow}{mlflow/\allowbreak mlflow} & \href{https://github.com/mlflow/mlflow/blob/master/LICENSE.txt}{Apache-2.0} & 1 & 6 & 1,417.3 \\*
\href{https://github.com/pyproj4/pyproj}{pyproj4/\allowbreak pyproj} & \href{https://github.com/pyproj4/pyproj/blob/main/LICENSE}{MIT} & 1 & 5 & 94.2 \\
\rowcolor{dividergray}
\multicolumn{5}{c}{\textbf{\textit{Desktop Environment}}} \\*
\href{https://github.com/materialsproject/pymatgen}{materialsproject/\allowbreak pymatgen} & \href{https://github.com/materialsproject/pymatgen/blob/main/LICENSE}{MIT} & 1 & 5 & 237.1 \\
\rowcolor{dividergray}
\multicolumn{5}{c}{\textbf{\textit{File Formats}}} \\*
\href{https://github.com/erocarrera/pefile}{erocarrera/\allowbreak pefile} & \href{https://github.com/erocarrera/pefile/blob/master/LICENSE}{MIT} & 1 & 4 & 6.6 \\*
\href{https://github.com/jmcnamara/XlsxWriter}{jmcnamara/\allowbreak XlsxWriter} & \href{https://github.com/jmcnamara/XlsxWriter/blob/main/LICENSE.txt}{BSD-2-Clause} & 1 & 5 & 33.8 \\*
\href{https://github.com/jupyter/nbconvert}{jupyter/\allowbreak nbconvert} & \href{https://github.com/jupyter/nbconvert/blob/main/LICENSE}{BSD-3-Clause} & 1 & 6 & 33.1 \\*
\href{https://github.com/python-excel/xlrd}{python-excel/\allowbreak xlrd} & \href{https://github.com/python-excel/xlrd/blob/master/LICENSE}{BSD-3-Clause / custom} & 1 & 5 & 9.4 \\
\rowcolor{dividergray}
\multicolumn{5}{c}{\textbf{\textit{Games/\allowbreak Entertainment}}} \\*
\href{https://github.com/rommapp/romm}{rommapp/\allowbreak romm} & \href{https://github.com/rommapp/romm/blob/master/LICENSE}{AGPL-3.0} & 1 & 5 & 233.9 \\*
\href{https://github.com/Tuxemon/Tuxemon}{Tuxemon/\allowbreak Tuxemon} & \href{https://github.com/Tuxemon/Tuxemon/blob/development/LICENSE}{GPL-3.0} & 1 & 6 & 127.0 \\
\rowcolor{dividergray}
\multicolumn{5}{c}{\textbf{\textit{Home Automation}}} \\*
\href{https://github.com/home-assistant/supervisor}{home-assistant/\allowbreak supervisor} & \href{https://github.com/home-assistant/supervisor/blob/main/LICENSE}{Apache-2.0} & 1 & 8 & 82.9 \\
\rowcolor{dividergray}
\multicolumn{5}{c}{\textbf{\textit{Internet}}} \\*
\href{https://github.com/boto/boto3}{boto/\allowbreak boto3} & \href{https://github.com/boto/boto3/blob/develop/LICENSE}{Apache-2.0} & 1 & 5 & 8.5 \\*
\href{https://github.com/camel-ai/camel}{camel-ai/\allowbreak camel} & \href{https://github.com/camel-ai/camel/blob/master/LICENSE}{Apache-2.0} & 4 & 25 & 160.2 \\*
\href{https://github.com/cherrypy/cherrypy}{cherrypy/\allowbreak cherrypy} & \href{https://github.com/cherrypy/cherrypy/blob/main/LICENSE.md}{BSD-3-Clause} & 1 & 6 & 17.3 \\*
\href{https://github.com/encode/httpcore}{encode/\allowbreak httpcore} & \href{https://github.com/encode/httpcore/blob/master/LICENSE.md}{BSD-3-Clause} & 1 & 8 & 8.2 \\*
\href{https://github.com/googleapis/google-api-python-client}{googleapis/\allowbreak google-api-python-client} & \href{https://github.com/googleapis/google-api-python-client/blob/main/LICENSE}{Apache-2.0} & 1 & 5 & 9.2 \\*
\href{https://github.com/pallets/flask}{pallets/\allowbreak flask} & \href{https://github.com/pallets/flask/blob/main/LICENSE.txt}{BSD-3-Clause} & 1 & 6 & 10.4 \\*
\href{https://github.com/python-restx/flask-restx}{python-restx/\allowbreak flask-restx} & \href{https://github.com/python-restx/flask-restx/blob/master/LICENSE}{BSD-3-Clause} & 1 & 5 & 13.7 \\
\rowcolor{dividergray}
\multicolumn{5}{c}{\textbf{\textit{Multimedia}}} \\*
\href{https://github.com/cuthbertLab/music21}{cuthbertLab/\allowbreak music21} & \href{https://github.com/cuthbertLab/music21/blob/master/LICENSE}{BSD-3-Clause} & 1 & 4 & 122.1 \\*
\href{https://github.com/deepchem/deepchem}{deepchem/\allowbreak deepchem} & \href{https://github.com/deepchem/deepchem/blob/master/LICENSE}{MIT} & 1 & 4 & 56.0 \\*
\href{https://github.com/harry0703/MoneyPrinterTurbo}{harry0703/\allowbreak MoneyPrinterTurbo} & \href{https://github.com/harry0703/MoneyPrinterTurbo/blob/main/LICENSE}{MIT} & 3 & 13 & 31.8 \\*
\href{https://github.com/Kometa-Team/Kometa}{Kometa-Team/\allowbreak Kometa} & \href{https://github.com/Kometa-Team/Kometa/blob/master/LICENSE}{MIT} & 1 & 5 & 31.5 \\*
\href{https://github.com/morpheus65535/bazarr}{morpheus65535/\allowbreak bazarr} & \href{https://github.com/morpheus65535/bazarr/blob/master/LICENSE}{GPL-3.0} & 1 & 4 & 993.8 \\
\rowcolor{dividergray}
\multicolumn{5}{c}{\textbf{\textit{Office/\allowbreak Business}}} \\*
\href{https://github.com/hummingbot/hummingbot}{hummingbot/\allowbreak hummingbot} & \href{https://github.com/hummingbot/hummingbot/blob/master/LICENSE}{Apache-2.0} & 1 & 5 & 242.7 \\*
\href{https://github.com/meta-pytorch/captum}{meta-pytorch/\allowbreak captum} & \href{https://github.com/meta-pytorch/captum/blob/master/LICENSE}{BSD-3-Clause} & 1 & 9 & 43.6 \\
\rowcolor{dividergray}
\multicolumn{5}{c}{\textbf{\textit{Scientific/\allowbreak Engineering}}} \\*
\href{https://github.com/GPflow/GPflow}{GPflow/\allowbreak GPflow} & \href{https://github.com/GPflow/GPflow/blob/develop/LICENSE}{Apache-2.0} & 1 & 5 & 10.4 \\*
\href{https://github.com/kubeflow/pipelines}{kubeflow/\allowbreak pipelines} & \href{https://github.com/kubeflow/pipelines/blob/master/LICENSE}{Apache-2.0} & 1 & 5 & 425.5 \\*
\href{https://github.com/NeuroTechX/moabb}{NeuroTechX/\allowbreak moabb} & \href{https://github.com/NeuroTechX/moabb/blob/develop/LICENSE}{BSD-3-Clause} & 2 & 9 & 64.6 \\*
\href{https://github.com/online-ml/river}{online-ml/\allowbreak river} & \href{https://github.com/online-ml/river/blob/main/LICENSE}{BSD-3-Clause} & 1 & 5 & 37.6 \\*
\href{https://github.com/PyDMD/PyDMD}{PyDMD/\allowbreak PyDMD} & \href{https://github.com/PyDMD/PyDMD/blob/master/LICENSE}{MIT} & 1 & 4 & 2.8 \\
\rowcolor{dividergray}
\multicolumn{5}{c}{\textbf{\textit{Security}}} \\*
\href{https://github.com/data-privacy-stack/presidio}{data-privacy-stack/\allowbreak presidio} & \href{https://github.com/data-privacy-stack/presidio/blob/main/LICENSE}{MIT} & 3 & 15 & 37.1 \\*
\href{https://github.com/demisto/content}{demisto/\allowbreak content} & \href{https://github.com/demisto/content/blob/master/LICENSE}{MIT} & 1 & 7 & 1,647.4 \\*
\href{https://github.com/domainaware/parsedmarc}{domainaware/\allowbreak parsedmarc} & \href{https://github.com/domainaware/parsedmarc/blob/master/LICENSE}{Apache-2.0} & 1 & 5 & 27.3 \\*
\href{https://github.com/oauthlib/oauthlib}{oauthlib/\allowbreak oauthlib} & \href{https://github.com/oauthlib/oauthlib/blob/master/LICENSE}{BSD-3-Clause} & 1 & 7 & 10.3 \\*
\href{https://github.com/ossf/cve-bin-tool}{ossf/\allowbreak cve-bin-tool} & \href{https://github.com/ossf/cve-bin-tool/blob/main/LICENSE.md}{GPL-3.0} & 1 & 4 & 47.5 \\*
\href{https://github.com/PurpleAILAB/Decepticon}{PurpleAILAB/\allowbreak Decepticon} & \href{https://github.com/BitterSecurity/Decepticon/blob/main/LICENSE}{Apache-2.0} & 1 & 4 & 123.5 \\
\rowcolor{dividergray}
\multicolumn{5}{c}{\textbf{\textit{Software Development}}} \\*
\href{https://github.com/arrow-py/arrow}{arrow-py/\allowbreak arrow} & \href{https://github.com/arrow-py/arrow/blob/master/LICENSE}{Apache-2.0} & 1 & 15 & 11.9 \\*
\href{https://github.com/ContextLab/hypertools}{ContextLab/\allowbreak hypertools} & \href{https://github.com/ContextLab/hypertools/blob/master/LICENSE}{MIT} & 1 & 4 & 2.0 \\*
\href{https://github.com/damnever/pigar}{damnever/\allowbreak pigar} & \href{https://github.com/damnever/pigar/blob/main/LICENSE}{BSD-3-Clause} & 1 & 4 & 1.9 \\*
\href{https://github.com/dask/distributed}{dask/\allowbreak distributed} & \href{https://github.com/dask/distributed/blob/main/LICENSE.txt}{BSD-3-Clause} & 1 & 4 & 96.4 \\*
\href{https://github.com/davidhalter/parso}{davidhalter/\allowbreak parso} & \href{https://github.com/davidhalter/parso/blob/master/LICENSE.txt}{MIT / PSF-2.0} & 1 & 5 & 9.4 \\*
\href{https://github.com/evennia/evennia}{evennia/\allowbreak evennia} & \href{https://github.com/evennia/evennia/blob/main/LICENSE.txt}{BSD-3-Clause} & 1 & 5 & 94.0 \\*
\href{https://github.com/facebookresearch/hydra}{facebookresearch/\allowbreak hydra} & \href{https://github.com/hydra-ecosystem/hydra/blob/main/LICENSE}{MIT} & 1 & 5 & 35.2 \\*
\href{https://github.com/flwrlabs/flower}{flwrlabs/\allowbreak flower} & \href{https://github.com/flwrlabs/flower/blob/main/LICENSE}{Apache-2.0} & 1 & 4 & 202.9 \\*
\href{https://github.com/inventree/InvenTree}{inventree/\allowbreak InvenTree} & \href{https://github.com/inventree/InvenTree/blob/master/LICENSE}{MIT} & 1 & 5 & 186.2 \\*
\href{https://github.com/jacebrowning/memegen}{jacebrowning/\allowbreak memegen} & \href{https://github.com/jacebrowning/memegen/blob/main/LICENSE.txt}{MIT} & 1 & 5 & 3.5 \\*
\href{https://github.com/jupyter/jupyter\_client}{jupyter/\allowbreak jupyter\_client} & \href{https://github.com/jupyter/jupyter\_client/blob/main/LICENSE}{BSD-3-Clause} & 1 & 6 & 8.7 \\*
\href{https://github.com/langchain-ai/langchain}{langchain-ai/\allowbreak langchain} & \href{https://github.com/langchain-ai/langchain/blob/master/LICENSE}{MIT} & 2 & 10 & 257.5 \\*
\href{https://github.com/pydantic/pydantic}{pydantic/\allowbreak pydantic} & \href{https://github.com/pydantic/pydantic/blob/main/LICENSE}{MIT} & 1 & 5 & 148.1 \\*
\href{https://github.com/UKGovernmentBEIS/inspect\_ai}{UKGovernmentBEIS/\allowbreak inspect\_ai} & \href{https://github.com/UKGovernmentBEIS/inspect\_ai/blob/main/LICENSE}{MIT} & 1 & 4 & 526.9 \\
\rowcolor{dividergray}
\multicolumn{5}{c}{\textbf{\textit{System}}} \\*
\href{https://github.com/saltstack/salt}{saltstack/\allowbreak salt} & \href{https://github.com/saltstack/salt/blob/master/LICENSE}{Apache-2.0} & 1 & 5 & 839.3 \\*
\href{https://github.com/siliconcompiler/siliconcompiler}{siliconcompiler/\allowbreak siliconcompiler} & \href{https://github.com/siliconcompiler/siliconcompiler/blob/main/LICENSE}{Apache-2.0} & 2 & 10 & 69.3 \\
\rowcolor{dividergray}
\multicolumn{5}{c}{\textbf{\textit{Terminals}}} \\*
\href{https://github.com/aws/aws-cli}{aws/\allowbreak aws-cli} & \href{https://github.com/aws/aws-cli/blob/v2/LICENSE.txt}{Apache-2.0} & 1 & 5 & 81.1 \\*
\href{https://github.com/Azure/azure-cli}{Azure/\allowbreak azure-cli} & \href{https://github.com/Azure/azure-cli/blob/dev/LICENSE}{MIT} & 1 & 4 & 1,966.1 \\*
\href{https://github.com/boto/botocore}{boto/\allowbreak botocore} & \href{https://github.com/boto/botocore/blob/develop/LICENSE.txt}{Apache-2.0} & 3 & 14 & 70.7 \\*
\href{https://github.com/Textualize/textual}{Textualize/\allowbreak textual} & \href{https://github.com/Textualize/textual/blob/main/LICENSE}{MIT} & 1 & 4 & 79.8 \\
\rowcolor{dividergray}
\multicolumn{5}{c}{\textbf{\textit{Text Processing}}} \\*
\href{https://github.com/nltk/nltk}{nltk/\allowbreak nltk} & \href{https://github.com/nltk/nltk/blob/develop/LICENSE.txt}{Apache-2.0} & 1 & 6 & 84.9 \\*
\href{https://github.com/pemistahl/lingua-py}{pemistahl/\allowbreak lingua-py} & \href{https://github.com/pemistahl/lingua-py/blob/main/LICENSE.txt}{Apache-2.0} & 1 & 4 & 5.3 \\
\end{longtable}
\endgroup

\end{document}